\documentclass[12pt,twocolumn]{aastex631}
\usepackage{color}
\usepackage{hyperref}
\usepackage{epstopdf}
\usepackage{graphicx}
\usepackage[FIGTOPCAP]{subfigure}
\usepackage{CJKutf8}
\usepackage{amsmath}
\usepackage{longtable}

\begin{document}
	\title{GRB 220426A: A Thermal Radiation-Dominated Gamma-Ray Burst}
	
	\correspondingauthor{Zhi-Ping Jin}
	\email{jin@pmo.ac.cn}
	
	\author[0000-0002-8385-7848]{Yun Wang}
	\affiliation{Key Laboratory of Dark Matter and Space Astronomy, Purple Mountain Observatory, Chinese Academy of Sciences, Nanjing 210034, China}
	\affiliation{School of Astronomy and Space Science, University of Science and Technology of China, Hefei, Anhui 230026, China}
	
	\author[0000-0001-6076-9522]{Tian-Ci Zheng}
	\affiliation{Key Laboratory of Dark Matter and Space Astronomy, Purple Mountain Observatory, Chinese Academy of Sciences, Nanjing 210034, China}
	\affiliation{School of Astronomy and Space Science, University of Science and Technology of China, Hefei, Anhui 230026, China}
	
	\author{Zhi-Ping Jin}
	\affiliation{Key Laboratory of Dark Matter and Space Astronomy, Purple Mountain Observatory, Chinese Academy of Sciences, Nanjing 210034, China}
	\affiliation{School of Astronomy and Space Science, University of Science and Technology of China, Hefei, Anhui 230026, China}
	
	\begin{abstract}
		The physical composition of the ejecta of gamma-ray bursts (GRBs) remains an open question.
		The radiation mechanism of the prompt gamma rays is also in debate.
		This problem can be solved for the bursts hosting distinct thermal radiation.  However, the events with dominant thermal spectral components are still rare.
		In this work, we focus on GRB 220426A, a recent event detected by Fermi-GBM. 
		The time-resolved and time-integrated data analyses yield very hard low-energy spectral indices and rather soft high-energy spectral indices.
		This means that the spectra of GRB 220426A are narrowly distributed.
		And the Bayesian inference results are in favor of the multicolor blackbody (mBB) model.
		The physical properties of the relativistic outflow are calculated.
		Assuming a redshift $z= 1.4$, the bulk Lorentz factors $\Gamma$ of the shells are found to be between $274_{-18}^{+24}$ and $827_{-71}^{+100}$, and the corresponding photosphere radii $R_{\rm ph}$ are in the range of  $1.83_{-0.50}^{+0.52} \times 10^{11}$ and $2.97_{-0.15}^{+0.14} \times 10^{12}$ cm. 
		Similar to GRB 090902B, the time-resolved properties of GRB 220426A  satisfy the observed $\Gamma-L$ and $E_p-L$ correlations, where $L$ is the luminosity of the prompt emission and $E_{p}$ is the spectral peak energy.
		
	\end{abstract}
	
	\section{Introduction} \label{sec:intro}
	The mechanism responsible for prompt emission in gamma-ray bursts (GRBs) has long been a mystery.
	Our understanding of prompt emission have been revolutionized in the past few decades with the development of observations.
	In the relativistic fireball model, thermal radiation is a natural explanation for the prompt emission \citep{goodman1986gamma}, 
	However, most of GRB spectra observed during the \emph{CGRO/BATSE} era are typically non-thermal, 
	and well fitted by an empirical smoothly joined broken power-law function (the so-called “Band” function; \citep{band1993batse}).
	A common interpretation of these non-thermal spectra was derived from the standard internal shock model \citep{rees1994unsteady}. 
	In spite of the fact that the Band function is only an empirical function, its arguments can be used to test emission mechanisms and underlying particle distributions.
	The low-energy spectral index obtained by the Band function can be used to determine whether the optically thin synchrotron limit \citep{preece1998synchrotron,preece2002consistency} has been exceeded.
	For GRBs exceed this limit {\citep{crider1997evolution,ghirlanda2003extremely}}, other components need to be considered, such as the thermal component, usually described as a Planck function.
	{Furthermore, for synchrotron or synchrotron-SSC models, the observed correlation between peak energy and luminosity cannot be explained without invoking additional assumptions \citep{golenetskii1983correlation,amati2002intrinsic,zhang2002analysis,lloyd2004kinetic}.}
	
	The existence of quasi-thermal components was confirmed by BATSE observation \citep{ryde2005thermal,ryde2009quasi}.
	{In the Fermi era, more observations discovered this potential component, such as GRB 100724B \citep{guiriec2011detection}, GRB 110721A \citep{axelsson2012grb110721a}, GRB 100507 \citep{ghirlanda2013photospheric},
		GRB 120323A \citep{guiriec2013evidence} and GRB 101219B \citep{larsson2015evidence}.}
	{In some of these bursts, the authors verified the existence of non-dominant thermal components through statistical methods.}
	{While the rest of bursts with dominant thermal components were relatively weak, and only part of blackbody spectrum could be observed.}
	{For the current observations, it is extremely rare that the thermal component is dominant and the complete blackbody spectrum be observed.}
	{Therefor, more direct evidence just like a ``smoking gun"} is the observation of GRB 090902B \citep{abdo2009fermi},
	a very bright gamma-ray burst with a narrow spectrum.
	{\cite{ryde2010identification} confirmed the thermal component in GRB 090902B, which can be fitted by a multicolor blackbody (mBB) model.}
	
	Recently, our briefing system, based on Fermi's online update catalog  \citep{gruber2014fermi,bhat2016third,von2014second,von2020fourth},
	reported a potential thermal radiation dominated sample GRB 220426A, similar to GRB 090902B.
	Further analyses showed that both its low and high energy spectral indices exceeded the typical values of normal GRB, which indicates a narrow spectrum.
	In this work, we employ the Bayesian inference  \citep{thrane2019introduction,van2021bayesian} for parameter estimation and model selection of the spectrum,
	and the results are in favor of the mBB model.
	We determine physical properties of the relativistic outflow based upon the identification of emissions from the photosphere, 
	such as the bulk Lorentz factor $\Gamma$ and the radius of the photosphere $R_{\rm ph}$.
	
	The paper is organized as follows:
	In Section \ref{sec:obs_ana}, we present observations and results of our analysis of GRB 220426A.
	In Section \ref{sec:2}, we further characterize GRB 220426A based on the results of the spectral and temporal analysis.
	In Section \ref{sec:3}, we calculated the physical parameters of the outflow based on the photospheric radiation and compared the correlations explained by the photospheric radiation.
	In Section \ref{sec:4}, we summarize our results with some discussion.
	
	\section{Observation and Data analysis}\label{sec:obs_ana}
	The Fermi-GBM team reports the detection of GRB 220426A (trigger 672648596/220426285) \citep{malacaria2022grb}.
	Based on the online updated  Fermi GBM Burst Catalog \citep{gruber2014fermi,bhat2016third,von2014second,von2020fourth},
	we developed a python GRB daily briefing system, and the indicators fed back by the system attracted our attention to this GRB event.
	
	We further analyze the GRB 220426A based on Fermi-GBM's observations.
	Fermi-GBM \citep{meegan2009fermi} consists of 12 sodium iodide (NaI) detectors and 2 bismuth germanate (BGO) detectors.
	The detectors were selected based on their pointing direction and count rate.
	Therefore a NaI detector (n0) and a BGO detector (b0) are used in our analysis.
	In this work, {\tt GBM Data Tools} \citep{GbmDataTools} is used to process the Fermi-GBM data, which is a pure Python tool and easy to use.
	In Figure \ref{fig:1} (a), we present the GBM light curves for several energy bands.
	We recalculated the $T_{90}$ \citep{koshut1996systematic} of the GBM n0 detector in the energy range of 50 - 300 keV,
	and used Bayesian block technique \citep{scargle2013studies} to determine the time interval of this burst (see Figure \ref{fig:1} (b)).
	
	\subsection{Spectral Analysis}\label{sec:spec_ana}
	We perform both time-integrated and time-resolved spectral analysis of GRB 220426A, and the specific time interval is shown in Table \ref{tab:tab1} and Table \ref{tab:tab2}.
	Following the procedure described in \cite{2022arXiv220502982W},
	we extract the source and background spectra, as well as the corresponding instrumental response files for each time interval. 
	The spectrum of GRBs can generally be fitted by a smoothly joined broken power-law function (the so-called “Band” function; \citealt{band1993batse}).
	The Band function is written as
	\begin{equation}
		N(E)=
		\begin{cases}
			A(\frac{E}{100\,{\rm keV}})^{\alpha}{\rm exp}{(-\frac{E}{E_0})}, \mbox{if $E<(\alpha-\beta)E_{0}$ }\\
			A[\frac{(\alpha-\beta)E_0}{100\,{\rm keV}}]^{(\alpha-\beta)}{\rm exp}{(\beta-\alpha)}(\frac{E}{100\,{\rm keV}})^{\beta},
			\mbox{if $E > (\alpha-\beta)E_{0}$}
		\end{cases}
		\label{eq:band}
	\end{equation}
	where \emph{A} is the normalization constant, \emph{E} is the energy in unit of keV, $\alpha$ is the low-energy photon spectral index, $\beta$ is the high-energy photon spectral index, and \emph{E$_{0}$} is the break energy in the spectrum.
	The peak energy in the $\nu F_\nu$ spectrum $E_{p}$ is equal to $E_{0}\times(2+\alpha)$.
	Additionally, if the count rate of high-energy photons is relatively low, the high-energy spectral index $\beta$ may not be constrained.
	The cutoff power-law function (CPL) can be used in this situation,
	\begin{equation}
		{ N(E)=A(\frac{E}{100\,{\rm keV}})^{\alpha}{\rm exp}(-\frac{E}{E_c}) },
	\end{equation}
	where \emph{$\alpha$} is the power law photon spectral index, \emph{E$_{c}$} is the break energy in the spectrum,
	and the peak energy $E_{p}$ is equal to $E_{c}\times(2+\alpha)$.
	When considering the thermal radiation component, the photon spectrum formula for blackbody radiation is usually expressed as
	\begin{equation}
		{ N(E)=\frac{ 8.0525\times K E^2}{(kT)^4 (e^{(E/kT)}-1)}}
	\end{equation}
	where \emph{kT} is the blackbody temperature keV; \emph{K} is the \emph{L$_{39}$}/ \emph{D$_{10}^2$}, where \emph{L$_{39}$} is the
	source luminosity in units of {10$^{39}$} erg/s and \emph{D$_{10}$} is the distance to the source in units of 10 kpc.
	Due to angle dependence of the Doppler shift, the observed blackbody temperature depends on the latitude angle. 
	Similarly to the optical depth, the photospheric radius increases with angle  \citep{pe2008temporal}. 
	It has a similar effect on outflow density profiles that are angle-dependent.
	The mBB is therefore a better description of the photospheric component than a single Planck function.
	By Considering the superposition of Planck functions at different temperatures, the phenomenological mBB model can be obtained \citep{ryde2010identification}. 
	The mBB model we use is modified by \cite{hou2018multicolor}, 
	\begin{equation}
		N(E)=\frac{8.0525(m+1)K}{\Big[\big(\frac{T_{\rm max}}{T_{\rm min}}\big)^{m+1}-1\Big]}\Big(\frac{kT_{\rm min}}{\rm keV}\Big)^{-2}I(E),\label{N(E)}
	\end{equation}
	where
	\begin{equation}
		I(E)=\Big(\frac{E}{kT_{\rm min}}\Big)^{m-1}\int_{\frac{E}{kT_{\rm max}}}^{\frac{E}{kT_{\rm min}}}\frac{x^{2-m}}{e^x-1}dx,\label{I(E)}
	\end{equation}
	where $x=E/kT$, the temperature range from $kT_{\rm min}$ to $kT_{\rm max}$,
	and the index $m$ of the temperature determines the shape of spectra.
	The mBB model approximates the spectrum of a pure blackbody when $m$ = 2.
	In addition, we also consider the case of each model plus a single power-law (PL) function with exponent $\gamma$.
	
	We employ the Bayesian inference \citep{thrane2019introduction,van2021bayesian} approach for parameter estimation and model selection by using the nested sampling algorithm Dynesty \citep{speagle2020dynesty,skilling2006nested,higson2019dynamic} in {\tt Bilby} \citep{ashton2019bilby}.
	The $pgstat$ statistic \footnote{https://heasarc.gsfc.nasa.gov/xanadu/xspec/manual/\\XSappendixStatistics.html} is used in Bayesian inference.
	For model selection, the Bayesian evidence ($\mathcal{Z}$) can be expressed as follows:
	\begin{equation}
		\mathcal{Z} = \int \mathcal{L}(d|\theta) \pi(\theta) d\theta, 
	\end{equation}
	The ratio of the $\mathcal{Z}$ for two different models is called as the Bayes factor (BF) and the logarithm of the BF reads
	\begin{equation}
		\ln\text{BF}^\text{A}_\text{B} = \ln({\cal Z}_\text{A}) - \ln({\cal Z}_\text{B}) .
		\label{eq:7}
	\end{equation}
	When $\ln{\rm BF} > 8$, we have the ``strong evidence'' in favor of one hypothesis over the other for selecting one that is statistically rigorous \citep{thrane2019introduction}.
	
	The posterior parameters and model selection of each model are shown in Table \ref{tab:tab1} and Table \ref{tab:tab2}, 
	and the evolution of the time-resolved spectrum is illustrated in Figure \ref{fig:par_lc}. 
	It is noteworthy that in the whole burst, the low-energy photon spectral index obtained by both the Band model and the CPL model are exceed the limit of the synchrotron shock model,
	also known as the ``Line of Death" \citep{preece1998synchrotron,preece2002consistency}.
	Furthermore, all high-energy photon spectral indexes are also exceed typical values ($\beta \sim$ -2) \citep{preece2000batse},
	which most likely correspond to the exponential decay of the Planck function at the highest temperature.
	In the fitting result of the time-integrated spectrum, the mBB+PL model obtained the highest evidence,
	and the observed photon count spectrum and $\nu F_{\nu}$ spectrum are shown in the upper panel of Figure \ref{fig:tot_mbb}.
	{The calculation results show that the quasi-thermal component (i.e. mBB) flux accounted for 86$\%$ of the total flux (in Fermi-GBM energy range, 8-40000 keV).}
	{This means that the spectrum of this burst is thermal dominated, which is also confirmed in the results of the model selection (see Table \ref{tab:tab1}).}
	In the analysis of time-resolved spectra, most time slices (9/14) in favor of the mBB model.  
	{And in the other five time slices, there is no ``strong evidence'' to rule out the mBB model}.
	Due to the low count rate of high-energy photons, an additional PL component is not required in the fitting of the time-resolved spectrum.
	The posterior of $m$ in the mBB model of the time slice [$T_0$ + 2.61, $T_0$ + 2.99 s] is 0.52$_{-0.11}^{+0.10}$, which is the slice closest to the blackbody spectrum, 
	and its observed photon count spectrum and $\nu F_{\nu}$ spectrum are shown in the lower panel of Figure \ref{fig:tot_mbb}.
	
	\section{Characteristics}\label{sec:2}
	\subsection{$E_{p,z}-E_{\gamma,\rm iso}$ correlation}\label{sec:amati}
	With the posterior parameters of the spectral analysis determined in Section \ref{sec:spec_ana},
	we compare GRB 220426A and 090902B in the $E_{p,z}-E_{\gamma,\rm iso}$ correlation \citep{amati2002intrinsic,zhang2009discerning}, see Figure \ref{fig:t90_dis} (a).
	The cosmological parameters are set to \emph{H$_{0}$} = $\rm 69.6 ~k ms^{-1}~Mpc^{-1}$, $\Omega_{\rm m}= 0.29$,
	and $\Omega_{\rm \Lambda}= 0.71$, to calculate the isotropic equivalent energy $E_{\gamma, {\rm iso}}$.
	We calculated $E_{\gamma, {\rm iso}}$ and $E_{p,z}$ in different redshifts (from 0.1 to 5) due to the lack of precise observations of redshifts.
	For comparison, we also plot the thermal radiation-dominated case GRB 090902B on this diagram.
	Obviously, they are all in the range of long GRBs.
	Based on the confidence interval of 1 $\sigma$ for the long burst in the $E_{p,z}-E_{\gamma,\rm iso}$ correlation, the redshift of GRB 220426A is estimated to be $1.40_{-0.38}^{+1.49}.$
	
	\subsection{$T_{90}$-related correlation and distributions}\label{}
	We examine some $T_{90}$-related correlation and distributions in order to determine the characteristics of GRB 220426A.
	For example, \cite{minaev2020p} proposed a classification scheme that combines the correlation of $E_{\gamma, {\rm iso}}$ and $E_{p,z}$, as well as the bimodal distribution of $T_{90}$.
	To characterize $E_{p,z}-E_{\gamma,\rm iso}$ correlation, $EH$ is proposed as
	\begin{equation}\label{key}
		EH = \dfrac{(E_{p,z}/100{\rm keV})}{(E_{\gamma,\rm iso}/10^{52}{\rm erg})^{0.4}}.
	\end{equation}
	The $T_{90,z}$-$EH$ trajectories calculated in different redshift (from 0.001 to 5) for GRB 220426A are shown in Figure \ref{fig:t90_dis} (b).
	In addition, the characteristics of $T_{90}$-hardness ratio (HR) and $T_{90}$-$E_p$ were also compared.
	The HR is calculated as the ratio of the observed counts in the range of 50 - 300 keV to the counts in the range of 10 - 50 keV \citep{goldstein2017ordinary}, and the $E_p$ of each burst is from the Fermi GBM Burst catalog \citep{gruber2014fermi,bhat2016third,von2014second,von2020fourth}.
	The $T_{90}$-HR and $T_{90}$-$E_p$ of GRB 220426A and GRB 090902B are plotted on Figure \ref{fig:t90_dis} (c) and (d) along with other catalog bursts, and the contour of the distribution was fitted with two-component Gaussian mixture model by {\tt scikit-learn}.
	The result show that GRB 220426A has a similar hardness ratio compared to GRB 090902B, but the latter has a higher $E_p$.
	
	\subsection{Spectral lag}\label{sec:spec_lag}
	In most GRBs, there is a lag between the different energy bands, called spectral lag.
	A cross-correlation function (CCF) can be used to quantify such an effect since pulse peaks at different energy bands are delayed.
	It is widely used to calculate spectral lag \citep{band1997gamma,ukwatta2010spectral}.
	CCF functions were calculated for GRB 220426A in different energy bands from $T_0$ - 1 to $T_0$ + 12 s (see the left of Figure \ref{fig:lag}),
	and the peak values of CCF were calculated via polynomial fitting.
	We can estimate the uncertainty of lags by using Monte Carlo simulations \citep{ukwatta2010spectral} (see the right of Figure \ref{fig:lag}).
	The spectral lag for 10 - 20 keV to 250 - 300 keV is 1.96 $\pm$ 0.07 s, and increases with increasing energy band.
	{The spectral lag increases with energy may be related to the spectral evolution \citep{lu2018comprehensive}.
		And it is more inclined to the long-burst population in the spectral delay classification \citep{bernardini2015comparing}.}
	
	\section{DERIVED PHYSICAL PARAMETERS and CORRELATIONS IN THE PHOTOSPHERIC RADIATION MODEL}\label{sec:3}
	By identifying the emission from the photosphere, we are able to determine physical properties of the relativistic outflow,
	such as the bulk Lorentz factor $\Gamma$ and photospheric radius $R_{\rm ph}$ \citep{pe2007new}.
	Due to the lack of exact redshift information, the redshift was {roughly} set to 1.4 based on the estimation in Section \ref{sec:amati},
	{and the redshift uncertainty is not considered in subsequent calculations.}
	Under the assumption that the radius of the photosphere $R_{\rm ph}$ is large the saturation radius $R_{\rm s}$, the Lorentz factor is calculated as
	\begin{eqnarray}
		\Gamma_{\rm ph}=[(1.06)(1+z)^{2}d_{L}\frac{Y\sigma_{\rm T}F^{\rm ob}}{2m_{p}c^{3}\Re}]^{1/4},
		\label{Lorentz}
	\end{eqnarray}
	where $d_L$ is the luminosity distance, $\sigma_{\rm T}$ is the Thomson scattering cross section, and $F^{ob}$ is the observed flux.
	We set $Y$ = 2 in our calculations, which is the ratio between the total fireball energy and the energy emitted in the gamma rays.
	$\Re$ is expressed as
	\begin{eqnarray}
		\Re=(\frac{F^{\rm ob}_{\rm thermal}}{\sigma T^{4}_{\rm max}})^{1/2},
		\label{}
	\end{eqnarray}
	where $F^{\rm ob}_{\rm thermal}$ is the thermal radiation flux. $\sigma$ is Stefan’s constant.
	The radius of the photosphere $R_{\rm ph}$ can be expressed as
	\begin{eqnarray}
		R_{\rm ph}= \frac{L_{\rm tot}\sigma_{\rm T}}{8\pi\Gamma^{3}_{\rm ph}m_{\rm p} c^{3}} ,
		\label{}
	\end{eqnarray}
	where $L_{\rm tot} = 4\pi d_L^2 Y F^{\rm ob}$ is the total luminosity.
	The evolution of the bulk Lorentz factor $\Gamma$ and the photosphere radius $R_{\rm ph}$ in the time-resolved spectrum is shown in the bottom panel of Figure \ref{fig:par_lc}.
	The calculated bulk Lorentz factors $\Gamma$ range from $274_{-18}^{+24}$ and $827_{-71}^{+100}$, 
	and the corresponding photosphere radius $R_{\rm ph}$ ranges from $1.83_{-0.50}^{+0.52} \times 10^{11}$ and $2.97_{-0.15}^{+0.14} \times 10^{12}$ cm.
	
	The photospheric radiation model may explain some of the observed correlations \citep{fan2012photospheric},
	and we compared GRB 220426A with GRB 090902B in the $\Gamma-L$ and $E_{p}-L$ correlations, see Figure \ref{fig:fan_2012}.
	We obtained correlations ${\rm log}\Gamma = 2.42_{-0.06}^{+0.06} +  0.28_{-0.06}^{+0.06}{\rm log}L $ and ${\rm log}E_{\rm p} = 2.47_{-0.05}^{+0.05} +  0.40_{-0.05}^{+0.04}{\rm log}L $ by fitting the filtered data \citep{lu2012lorentz,zhang2012revisiting}.
	There is one obvious outlier (the first time slice) which is likely due to the way we calculate the bulk Lorentz factor, which is only valid when $R_{\rm ph} > R_{\rm s}$.
	With the first time slice excluded, 
	the time-resolved spectral of GRB 220426A and GRB 090902B are consistent with the two correlations ($\Gamma-L$ and $E_p-L$) that can be explained by the photospheric radiation model.
	
	\section{SUMMARY and Discussion}\label{sec:4}
	{For GRBs with very hard low-energy spectral indices and very soft high-energy spectral indices, 
		the most natural explanations are the temperature distribution from the mBB model and the exponential decay of the Planck function at the highest temperature.
		Using Bayesian inference, we confirmed that a mBB model was more appropriate for describing the spectrum of GRB 220426A, similar to GRB 090902B.}
	In addition, we also have carried out a detailed analysis of GRB 220426A, which can be summarized as follows:
	\begin{itemize}
		\item  In either time-integrated or time-resolved spectrum analysis, the low-energy spectral index exceeds the ``Line of Death" of synchrotron radiation,
		while the high-energy spectral index exceeds the typical value ($\beta \sim -2$).
		It means that GRB 220426A has the same narrow spectrum as GRB 090902B.
		\item GRB 220426A and GRB 09092B are consistent in $E_{\gamma, {\rm iso}}$ - $E_{p,z}$ correlation, $T_{90,z}$ - $EH$ correlations and $T_{90}$-related distributions, both are long GRBs with $E_p$ of several hundred keV.
		\item The temporal analysis indicates that GRB 220426A has obvious spectral lags that increase with increasing energy band.
		\item The bulk Lorentz factor $\Gamma$ is between $274_{-18}^{+24}$ and $827_{-71}^{+100}$, and the corresponding photosphere radius $R_{\rm ph}$ is between   $1.83_{-0.50}^{+0.52} \times 10^{11}$ and $2.97_{-0.15}^{+0.14} \times 10^{12}$ cm determined by the photosphere emission.
		\item The time-resolved spectrum of GRB 220426A and GRB 090902B are consistent with the two correlations ($\Gamma-L$ and $E_p-L$) that can be explained by the photospheric radiation model (see \citet{fan2012photospheric} for the details).
	\end{itemize}
	
	{According to the current research, the prompt emission spectrum of GRBs usually has three elemental spectral components \citep{zhang2011comprehensive}, namely, a non-thermal Band component, a quasi-thermal component, and another non-thermal power-law component extending to high energies.}
	{It is widely believed that the non-thermal Band component originates from the optically thin synchrotron radiation, while the quasi-thermal components originates from the photosphere.}
	{Most of the GRBs are dominated by non-thermal Band components, which are represented by} GRB 080916C \citep{abdo2009fermi}, 
	which has a series of time-resolved standard Band spectrum covering 6–7 orders of magnitude interpreted as initially magnetically dominated outflow \citep{zhang2009evidence}.
	{Nonetheless, some studies discuss the natural presence of photospheric thermal radiation component in GRBs \citep{meszaros2000steep,meszaros2002x,daigne2002expected,rees2005dissipative}.
		In addition, there is a lot of indirect evidence that thermal component is present in most GRBs, even close to 100 $\%$ (see \citet{2017IJMPD..2630018P} for the details).}	
	{For the case when the thermal component is not dominant, a detailed statistical analysis is made based on the existing data \citep{li2019thermal,li2020thermal}.
		In the analysis of some multi-pulse GRBs, the thermal component can be found in the duration of the burst, and is more common in the early phase \citep{li2021bayesian}.
		The representative of this evolution of jet composition from fireball to Poynting flux-dominated outflow is GRB 160625B \citep{zhang2018transition}.}
	
	{The thermal component is an inherent part of the cosmological fireball model, and it was expected early on \citep{goodman1986gamma,paczynski1986gamma}.
		{Such thermal components have intriguing implication on the initial outflow or the energy dissipation in the inner region. Likely, the GRB ejecta was launched via the neutrino and anti-neutrino annihilation, for which the initial outflow was an extremely hot baryonic fireball and a fraction of the thermal energy will be radiated directly \citep{piran1993hydrodynamics,1993ApJ...415..181M}.}
		{This can only happen for an extremely-high rate of the accretion of the material onto the rapidly rotating black hole otherwise the annihilation luminosity is not high enough \citep{zalamea2011neutrino,fan2011short}}.
		{If the GRB ejecta was mainly launched via the magnetic process (for instance, the BZ mechanism \citep{blandford1977electromagnetic}) and hence Poynting flux dominated, a distinct thermal radiation component appears if the magnetic reconnection takes place efficiently when the outflow was still optically thick.}
		{Since thermal radiation dominant GRBs are rare, the former scenario (i.e., an extremely high accretion rate) may be favored or alternatively the magnetic energy dissipation can only be efficient in some stringent constraints that need to be better understood.}
	
	GRB 220426A as such a rare sample has a dominant quasi-thermal component with a subdominant non-thermal power-law component, and its quasi-thermal component flux accounts for 86$\%$ of the total flux.
	In addition to the initial thermal photons from the fireball, it may also come from the friction between the jet components, or the jet components and the surrounding material \citep{beloborodov2010collisional,vurm2011gamma}.
	According to the analysis in the Section \ref{sec:2}, GRB 220426A is a long burst may originate from the core collapse model \citep{woosley1993gamma,paczynski1998gamma,fryer1999formation,macfadyen1999collapsars,popham1999hyperaccreting,woosley2006supernova}, and its jet will drill out of the collapsed material \citep{aloy2000relativistic,macfadyen2001supernovae}.
	Shock waves from the friction of the jet and the  stellar envelope will heat the plasma, and when this occurs below the photosphere, thermal photons are produced \citep{lazzati2009very,morsony2010origin}.
	In such a scenario, the predicted thermal emission time is related to the collapse time of stellar material \citep{aloy2000relativistic,morsony2007temporal,bromberg2011propagation}. 
	
{In addition, our work verifies the existence of photospheric radiation in GRB, and in very rare cases, it can even dominate the radiation. 
	The conditions under which photospheric radiation dominates are yet to be discovered. 
	In the analysis of the prompt emission spectrum of GRBs in the future, considering the significance of thermal components may become a paradigmatic analysis method.
	The Lorentz factor obtained from the thermal component conforms to the statistical relationship of the Lorentz factor calculated by other methods, which means that it is reliable to limit the physical properties of GRBs.
	In the time-resolved spectral analysis of GRB 220426A, the exact non-thermal radiation evolution cannot be given due to the low flux of high-energy photons. 
	It is expected that instruments with higher sensitivity and wider energy range (for example VLAST; \cite{fan2022very}) may solve this problem in the future, and in the case of high-confidence thermal components, the evolution of some physical properties of GRBs, such as photosphere radius and energy dissipation radius, can be more accurately constrainted.}

\section*{Acknowledgments}
We thank the anonymous referee for their helpful suggestions. 
We appreciate Yi-Zhong Fan and Fu-Wen Zhang for their important help in this work.
We acknowledge the use of the Fermi archive's public data.
This work is supported by NSFC under grant No. 11921003 and 11933010.

\software{\texttt{Matplotlib} \citep{Hunter:2007}, \texttt{Numpy} \citep{harris2020array}, \texttt{scikit-learn} \citep{scikit-learn},
	\texttt{bilby} \citep{ashton2019bilby}, \texttt{GBM Data Tools} \citep{GbmDataTools}}

\bibliography{bibtex}
\begin{figure}
\centering
\includegraphics[width=0.4\textwidth]{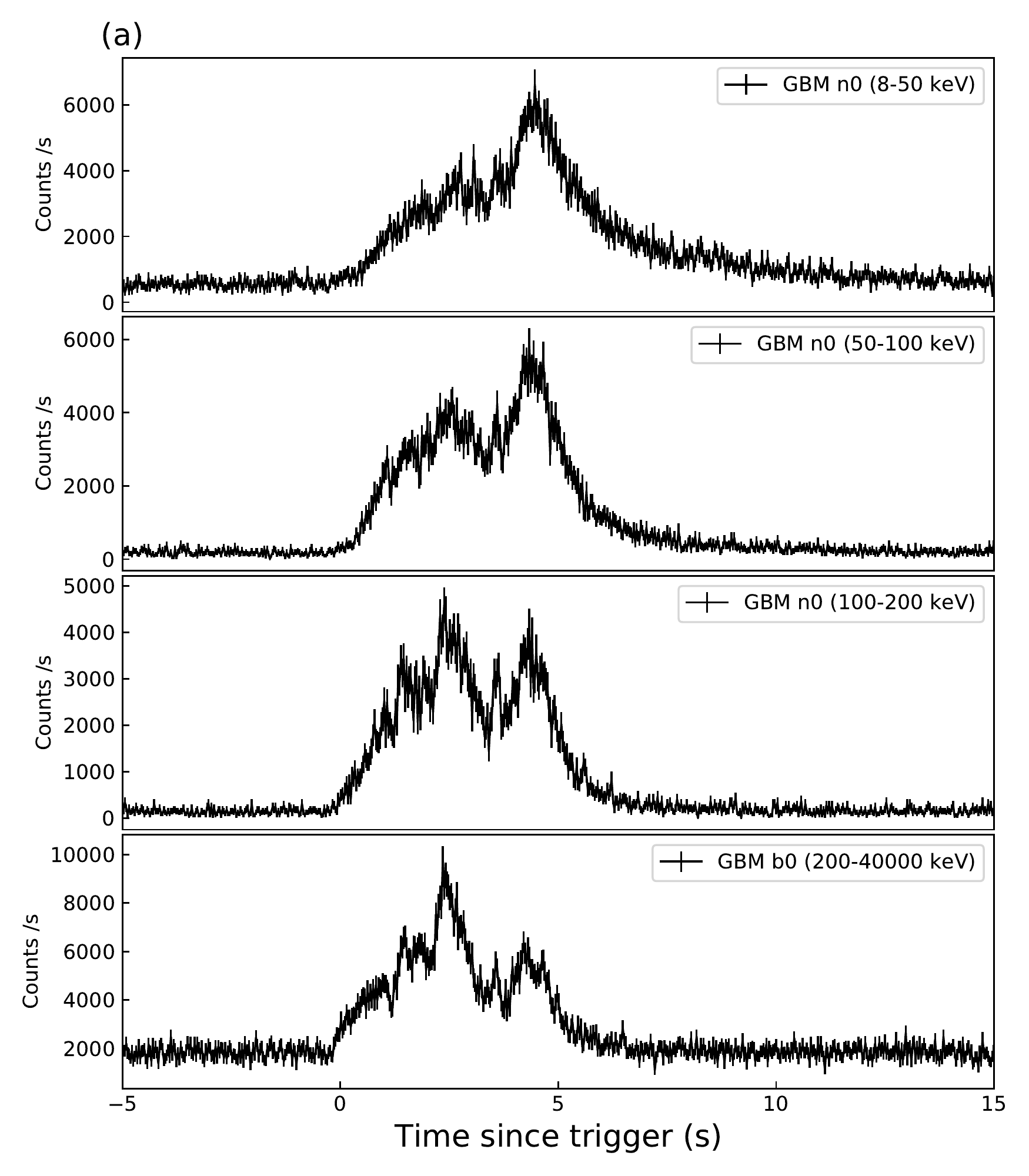}
\includegraphics[width=0.4\textwidth]{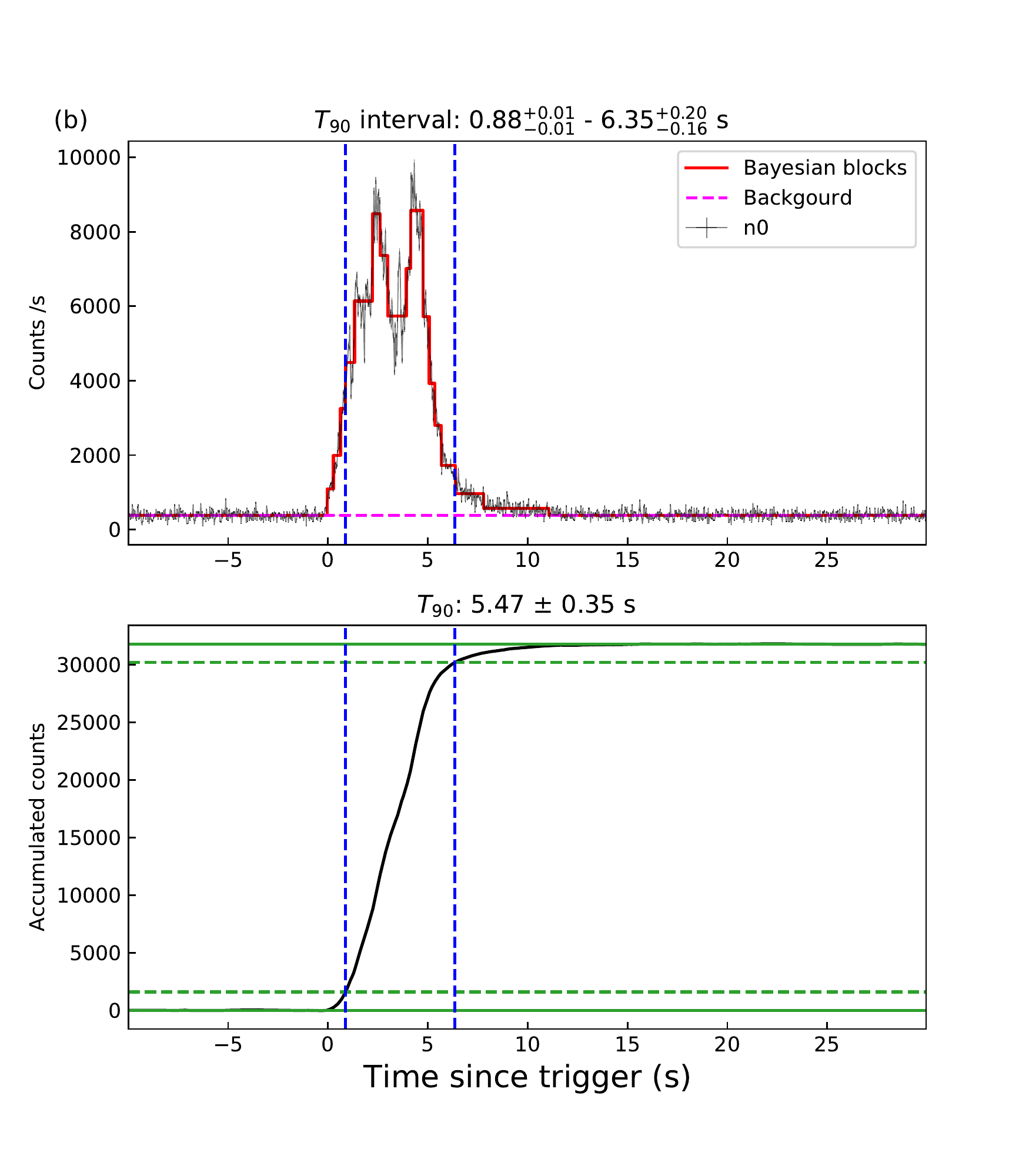}
\caption{Observational of GRB 220426A. (a) is the light curve of each energy band in different detectors.
	The binsize is set to 32 ms.
	(b) is the $T_{90}$ calculation.
	Upper panel of (b) shows the light curve of the NaI detector with an energy range of 50 - 300 keV with the bin size of 64 ms.
	The solid red line and dashed magenta line are the Bayesian block and the background, respectively.
	The bottom panel of (b) show the photon count accumulation curve.
	The green dashed line represents the range of values between 5$\%$ and 95$\%$ of the cumulative photon count.}
\label{fig:1}
\end{figure}
\begin{figure}
\centering
\includegraphics[width=0.5\textwidth]{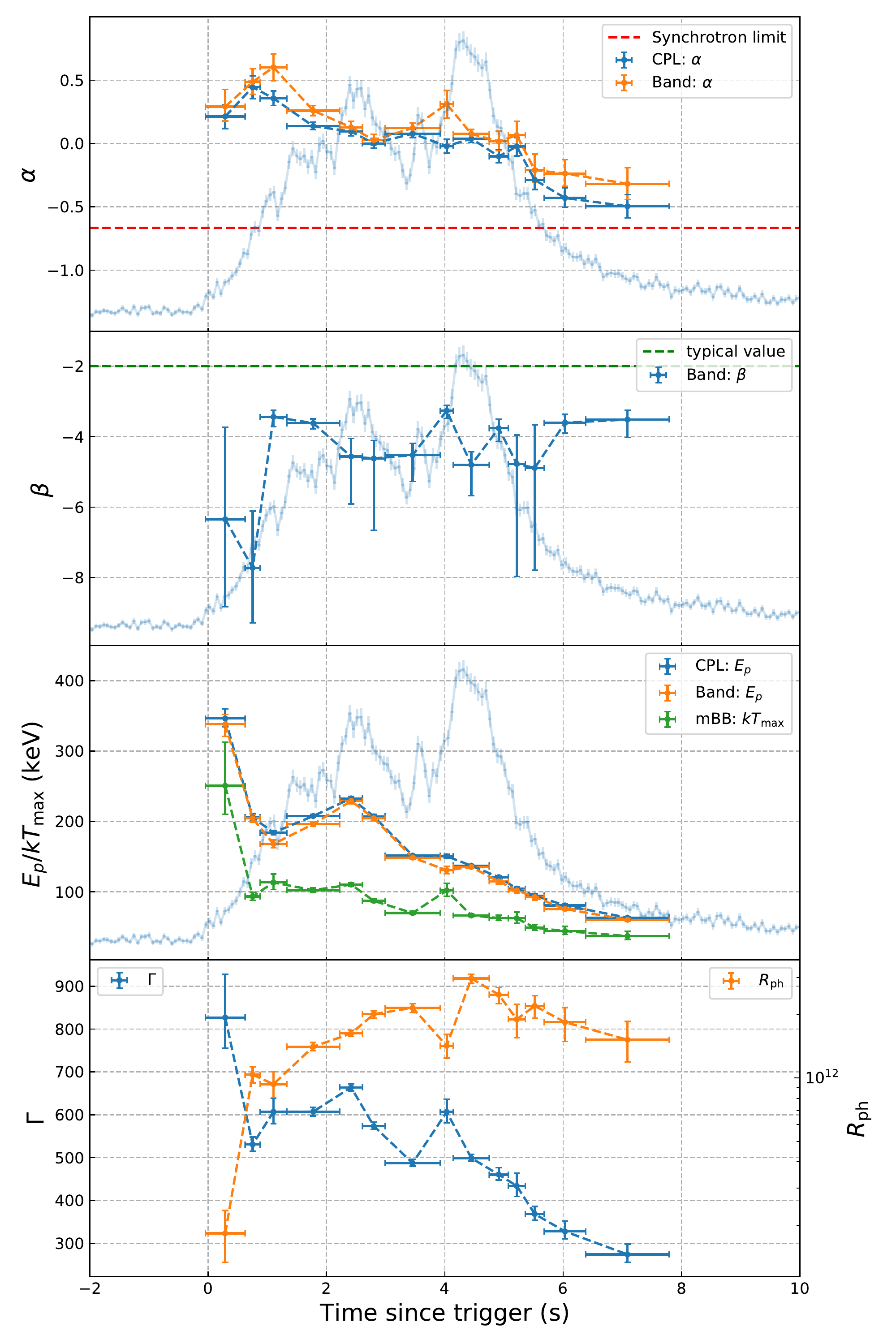}
\caption{The first three panels from top to bottom are the evolution diagrams of the posterior parameters of different models.
	The red dashed line in the first panel is the limit of the synchrotron shock model,
	also known as the ``Line of Death" \citep{preece1998synchrotron,preece2002consistency}.
	The green dashed line in the second panel represents the typical value of high-energy photon spectral index ($\beta \sim$ 2) \citep{preece2000batse}.
	The bottom panel is the evolution of the bulk Lorentz factor $\Gamma$ and the radius of the photosphere $R_{\rm ph}$.}
\label{fig:par_lc}
\end{figure}
\begin{figure}
\centering
\includegraphics[width=0.3\textwidth]{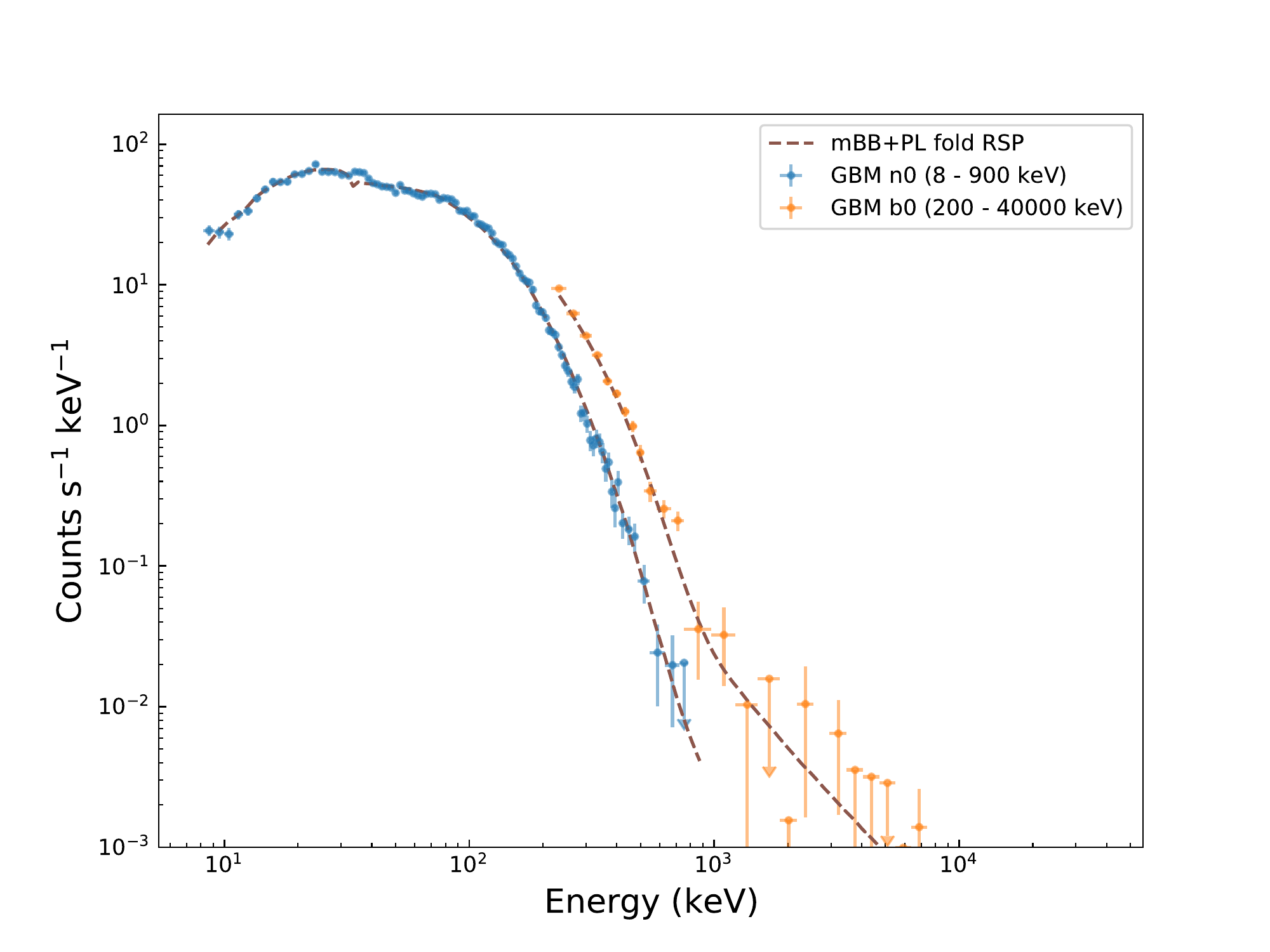}
\includegraphics[width=0.3\textwidth]{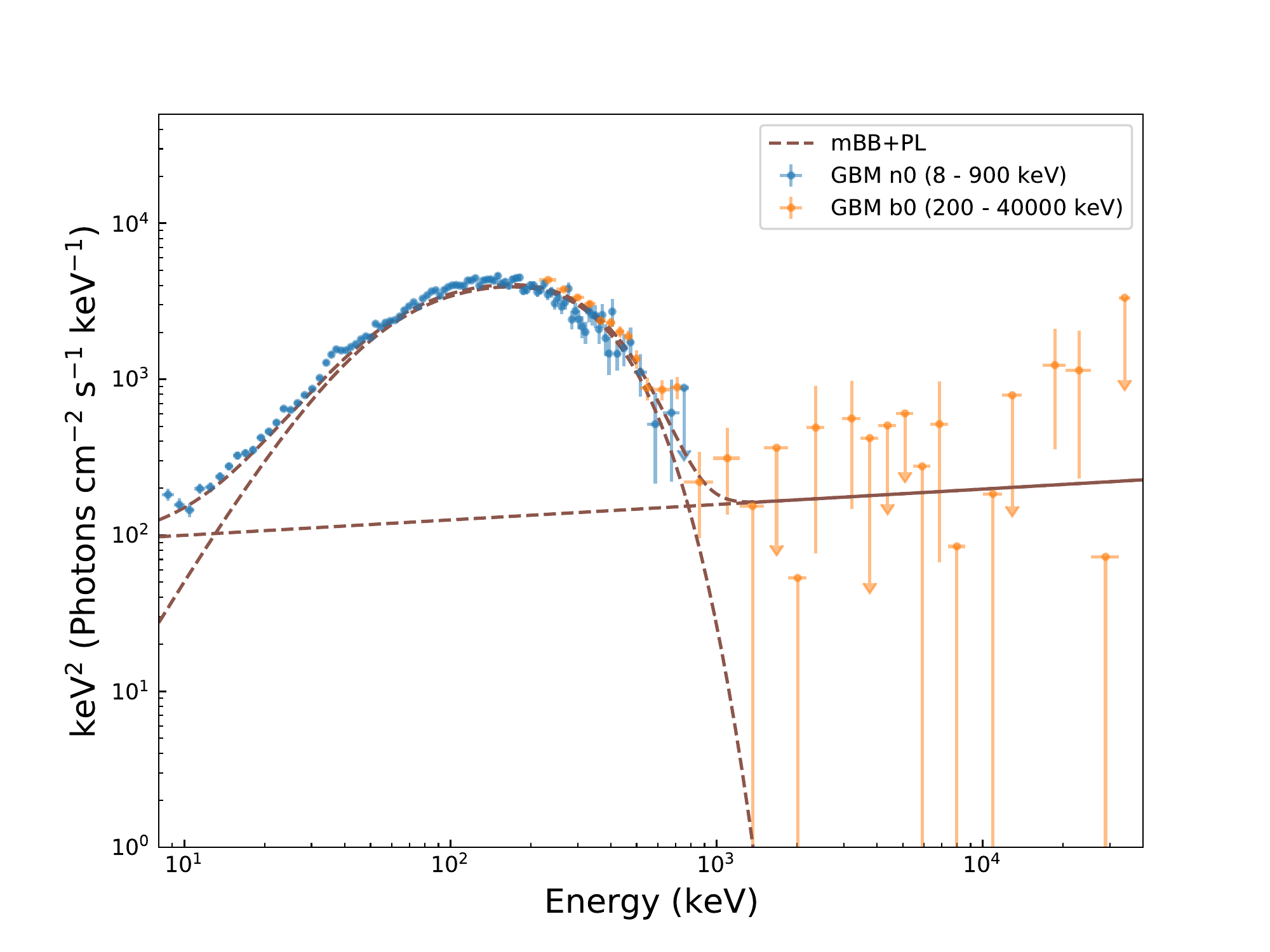}
\includegraphics[width=0.3\textwidth]{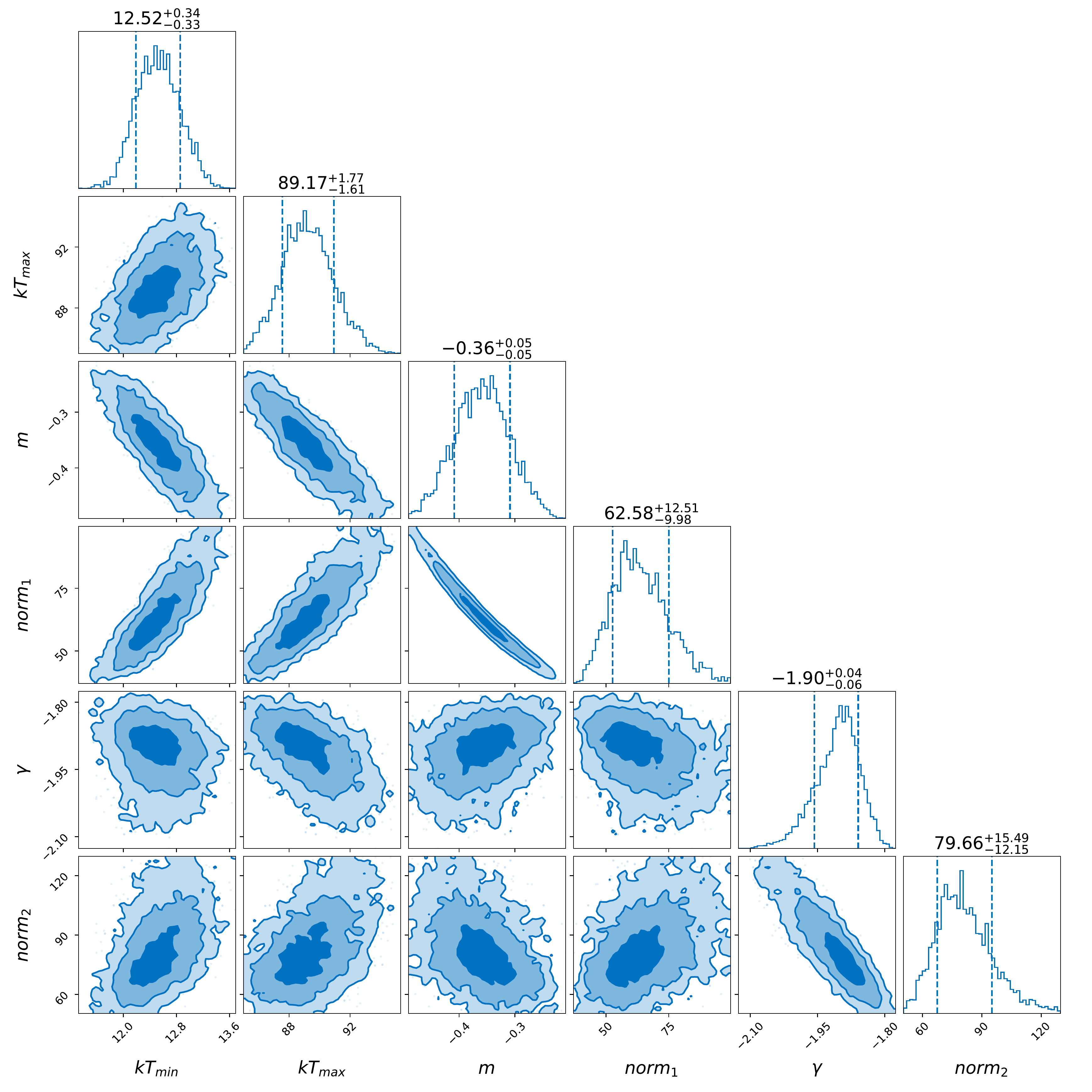}
\includegraphics[width=0.3\textwidth]{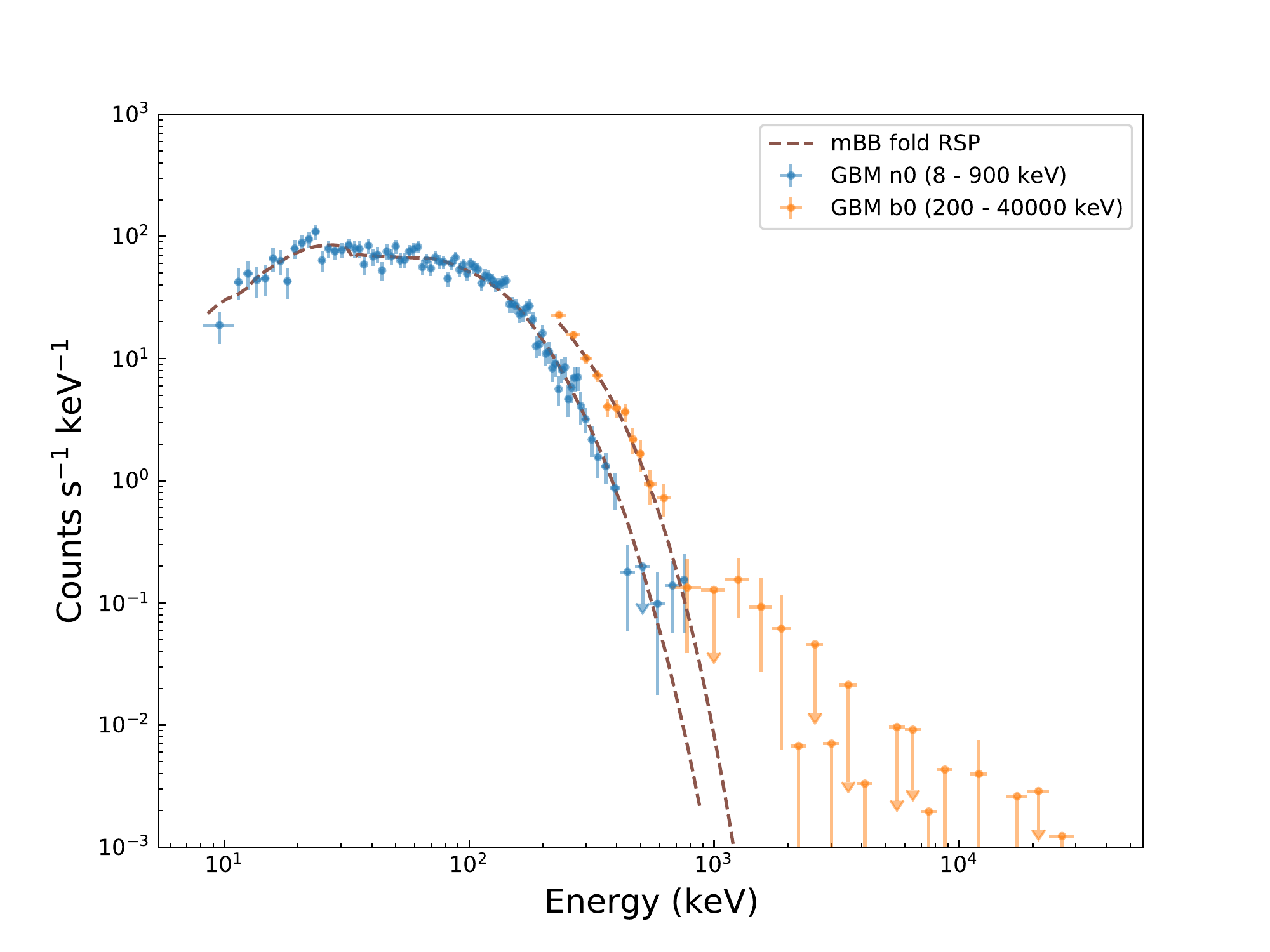}
\includegraphics[width=0.3\textwidth]{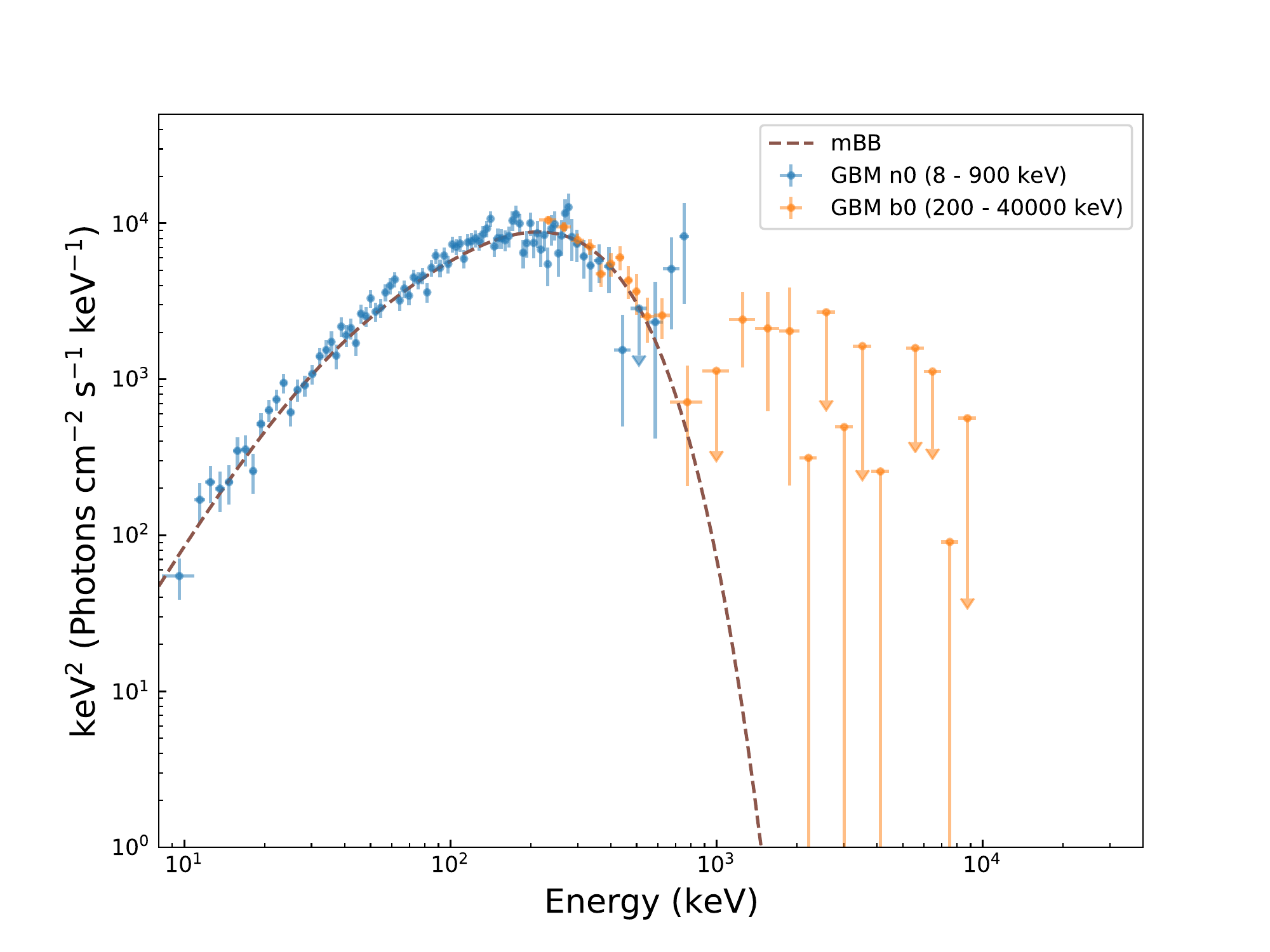}
\includegraphics[width=0.3\textwidth]{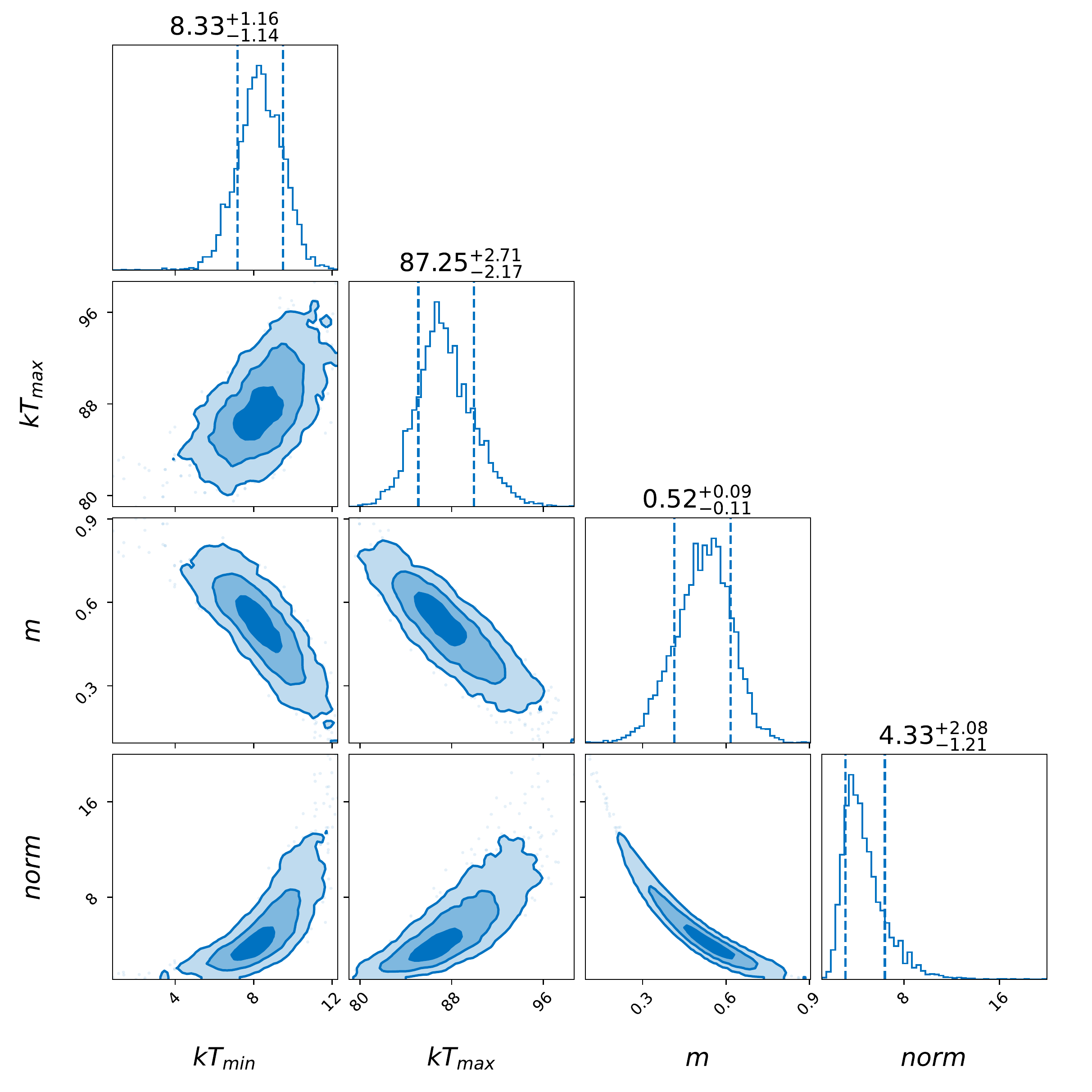}
\caption{Upper panel: Spectral analysis results of time-integrated spectrum ([$T_0$ - 0.05, $T_0$ + 7.79 s]).
	Upper left: the observed photon count spectra (scatter of different colors) and the photon spectra (brown dashed line) obtained by the mBB+PL model folded response file of each detector.
	Upper middle: the corresponding $\nu F_{\nu}$ spectra.
	Upper right: The posterior parameter distribution of the mBB+PL model.
	Lower panel: Spectral analysis results of a time-resolved spectrum slice ([$T_0$ + 2.61, $T_0$ + 2.99 s]). The description is consistent with Upper panel except that the model is a single mBB.}
\label{fig:tot_mbb}
\end{figure}
\begin{figure}
\centering
\includegraphics[width=0.45\textwidth]{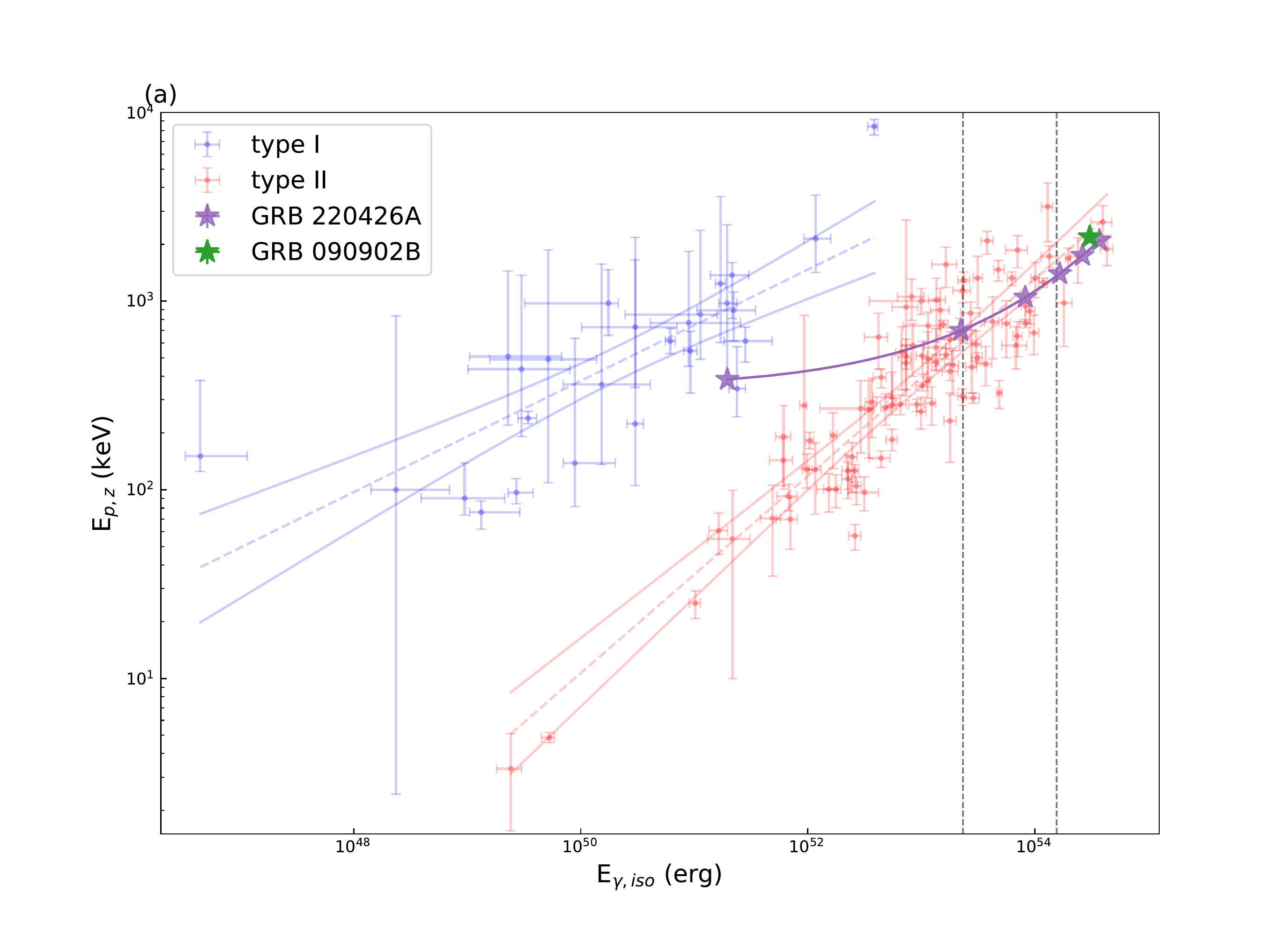}
\includegraphics[width=0.45\textwidth]{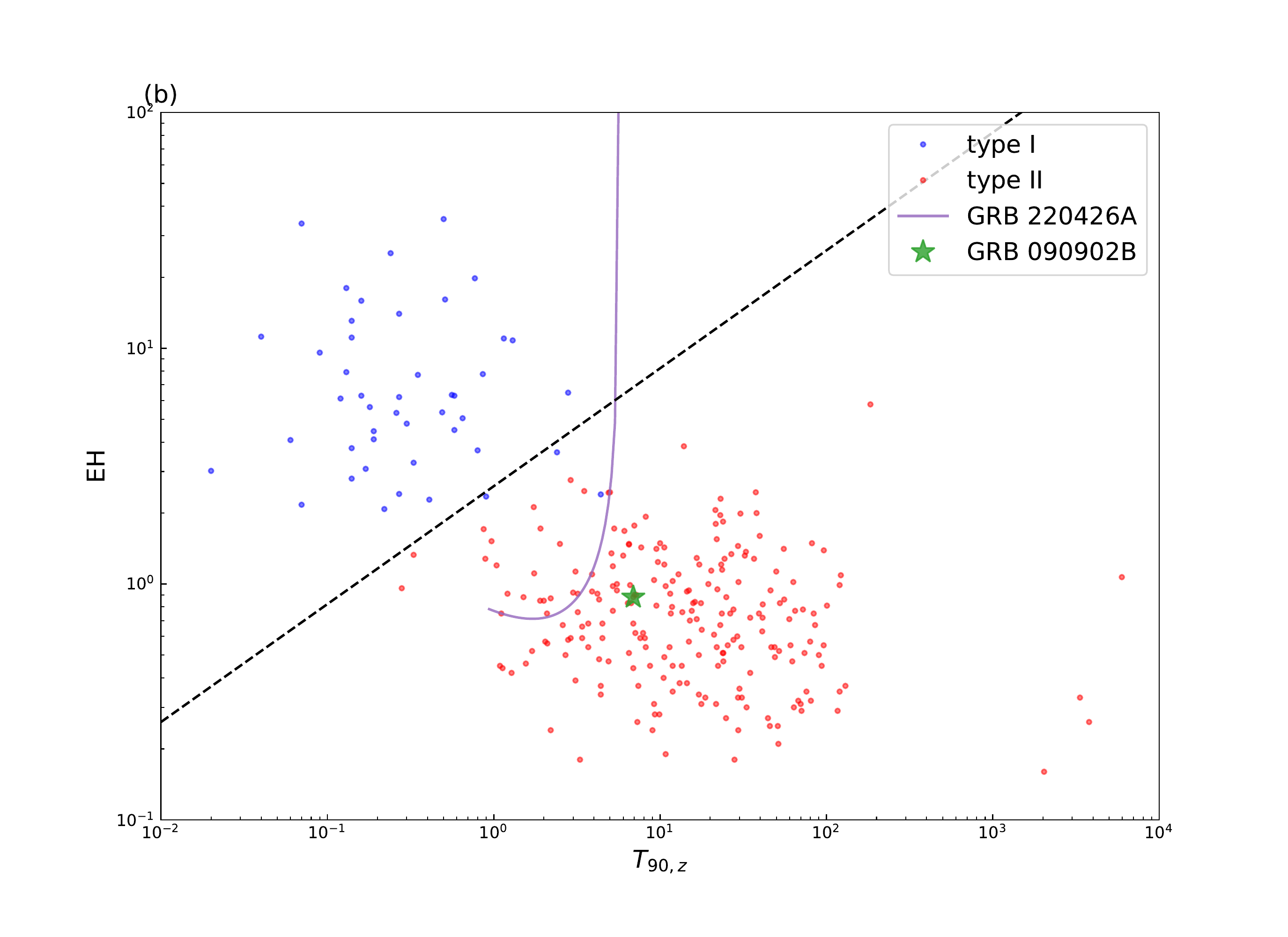}
\includegraphics[width=0.45\textwidth]{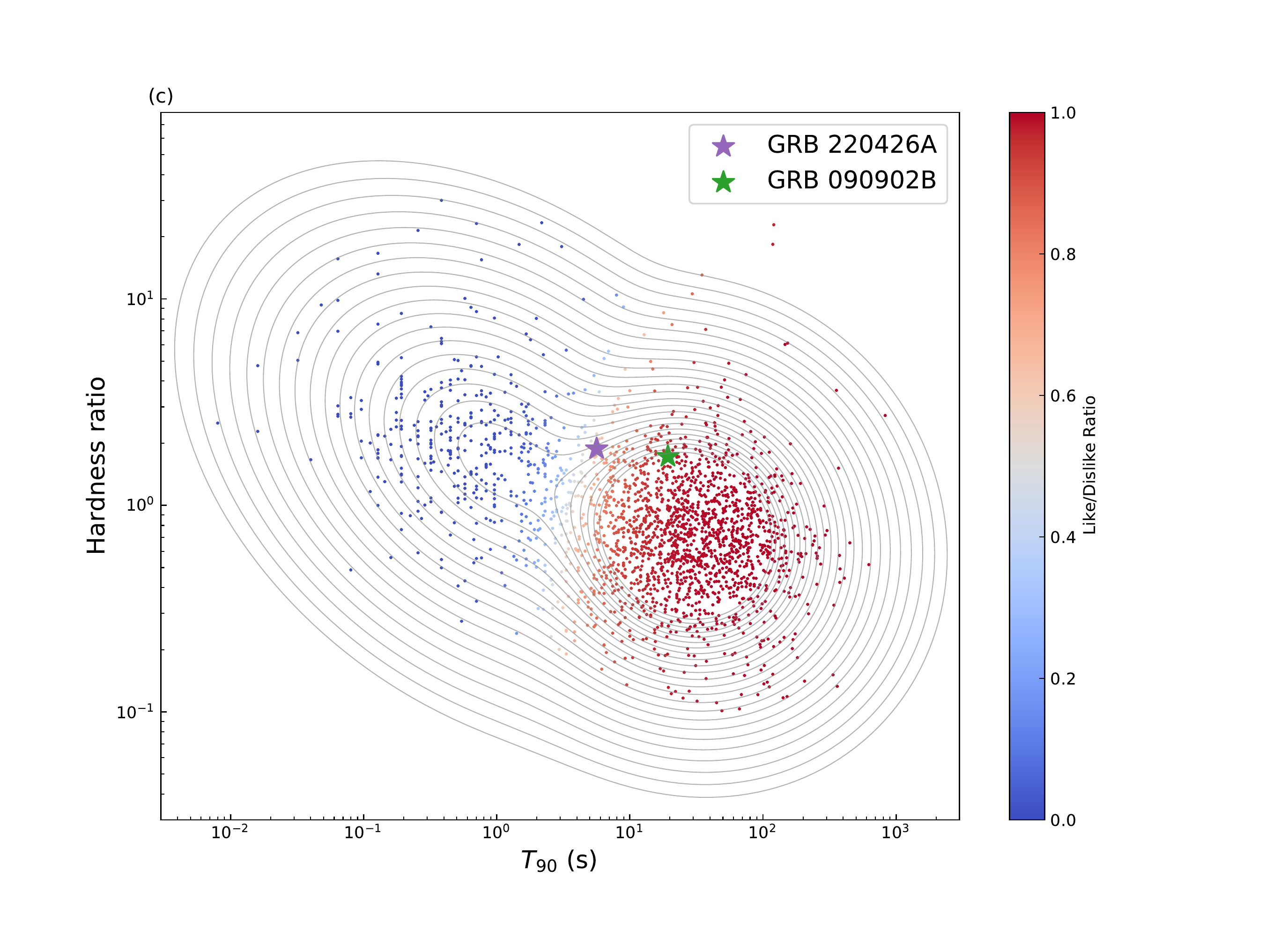}
\includegraphics[width=0.45\textwidth]{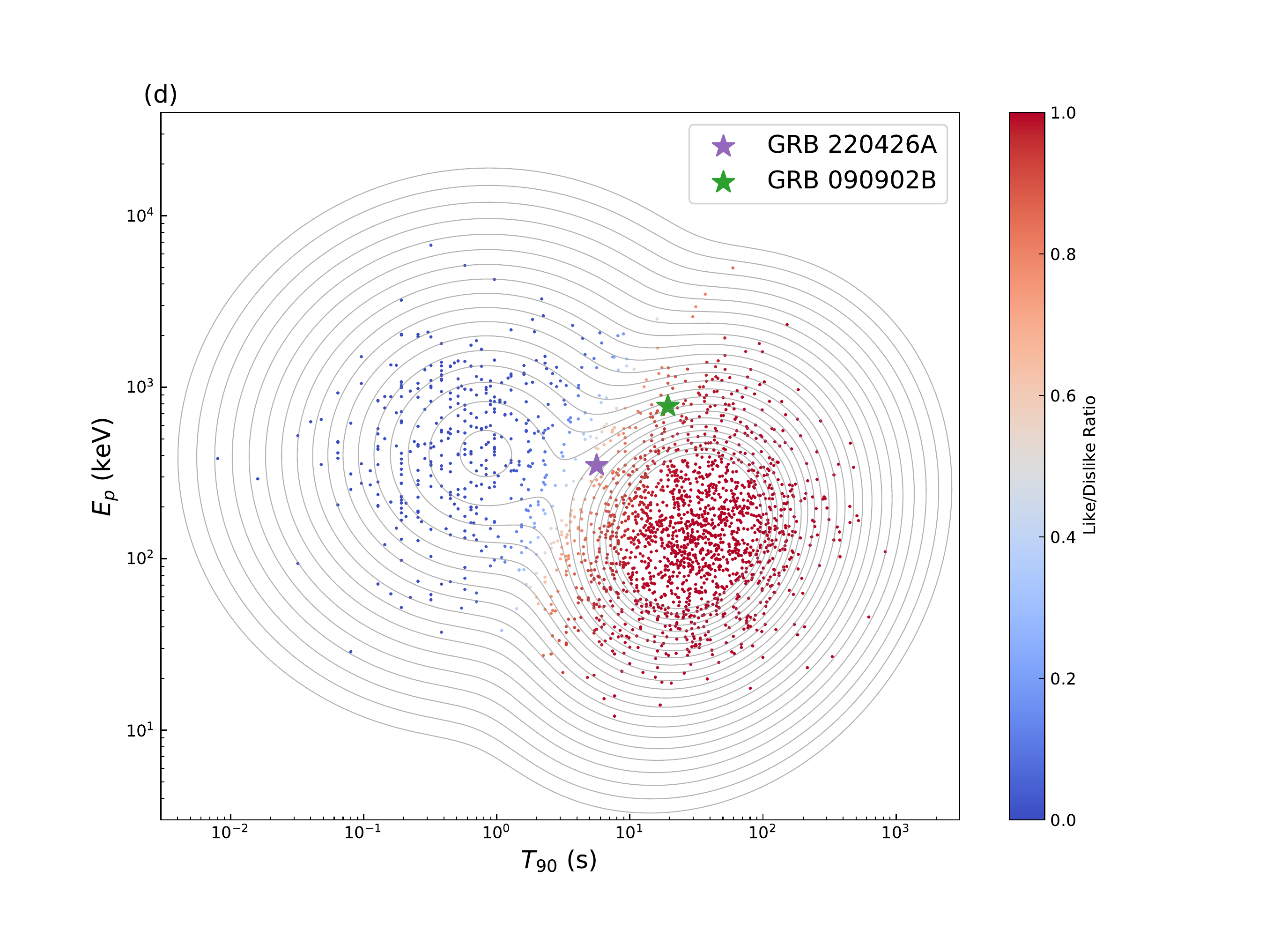}
\caption{Correlations and statistical distributions. 
	For comparison, purple and green represent GRB 220426A and GRB 090902B, respectively.
	The black dashed line is the range of the 1 $\sigma$ confidence interval of GRB 220426A in the $E_{\gamma, {\rm iso}}$ - $E_{p,z}$ correlation.
	(a) is the intrinsic spectral peak energy ($E_{p,z}$) and isotropic equivalent gamma-ray radiation energy ($E_{\gamma,iso}$) correlation diagram.
	Blue and red dashed lines represent the best-fit correlations for the type-I and type-II GRB samples, respectively.
	The trajectory indicate GRB 220426A in different redshift values (from 0.1 to 5).
	(b) is the $T_{90,z}$ - $EH$ diagram, and the rajectories is calculated at different redshift from 0.001 to 5.
	(c) and (d) give the $E_p$ and HR , respectively, and the comparison with other GRBs in the Fermi-GBM catalog.}
\label{fig:t90_dis}
\end{figure}
\begin{figure}
\centering
\includegraphics[width=0.45\textwidth]{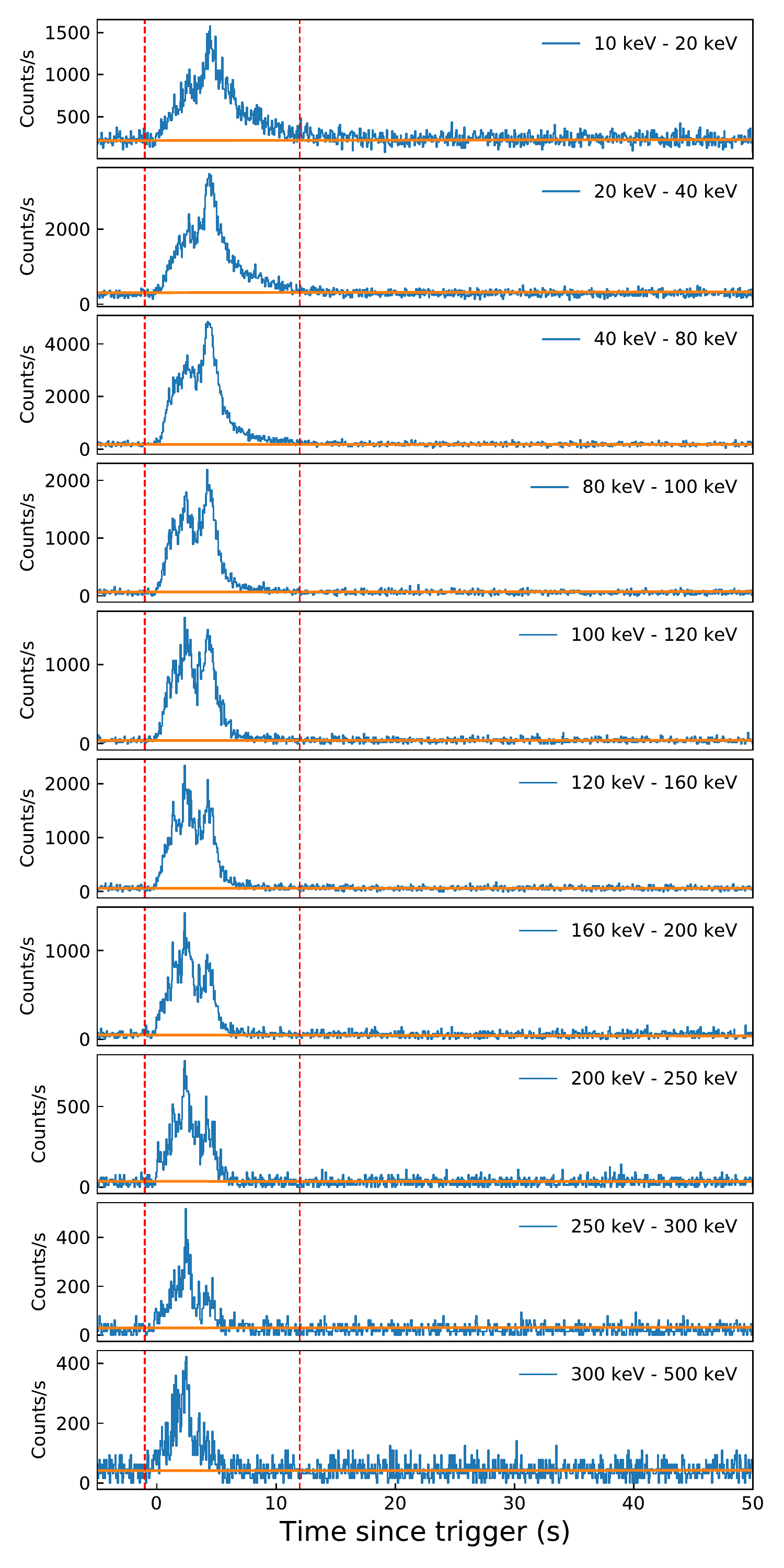}
\includegraphics[width=0.45\textwidth]{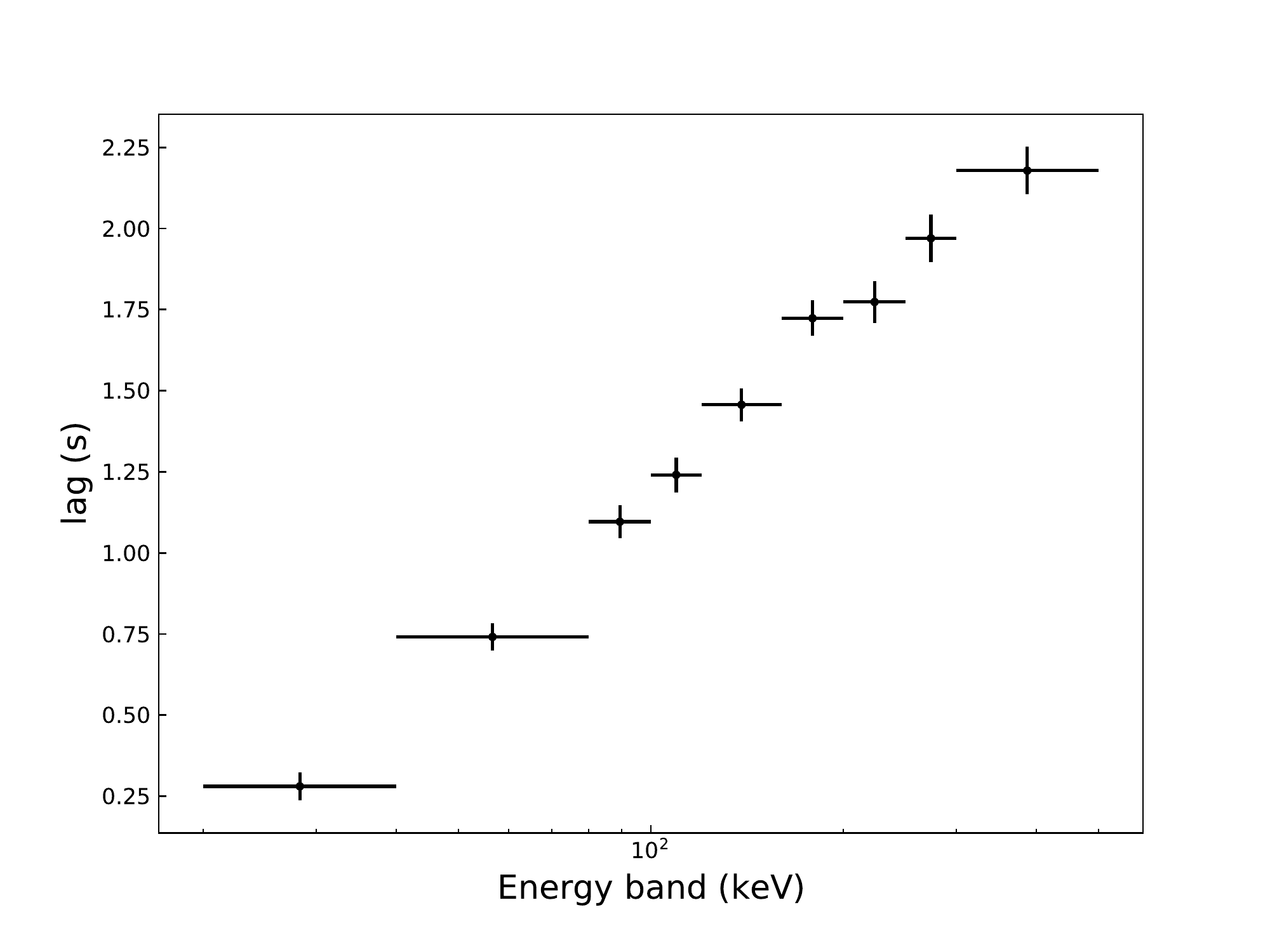}
\caption{Calculation of spectral lags in multiband light curves. 
	Left panel: the multiband light curve is from the Fermi-GBM n0 detector, and the interval in red dashed line are used to calculate the spectral lags.
	Right panel: the spectral lags between high energy bands and the lowest energy band (10 - 20 keV), 1 $\sigma$ error bars calculated from Monte Carlo simulations.}
\label{fig:lag}
\end{figure}
\begin{figure}
\centering
\includegraphics[width=0.6\textwidth]{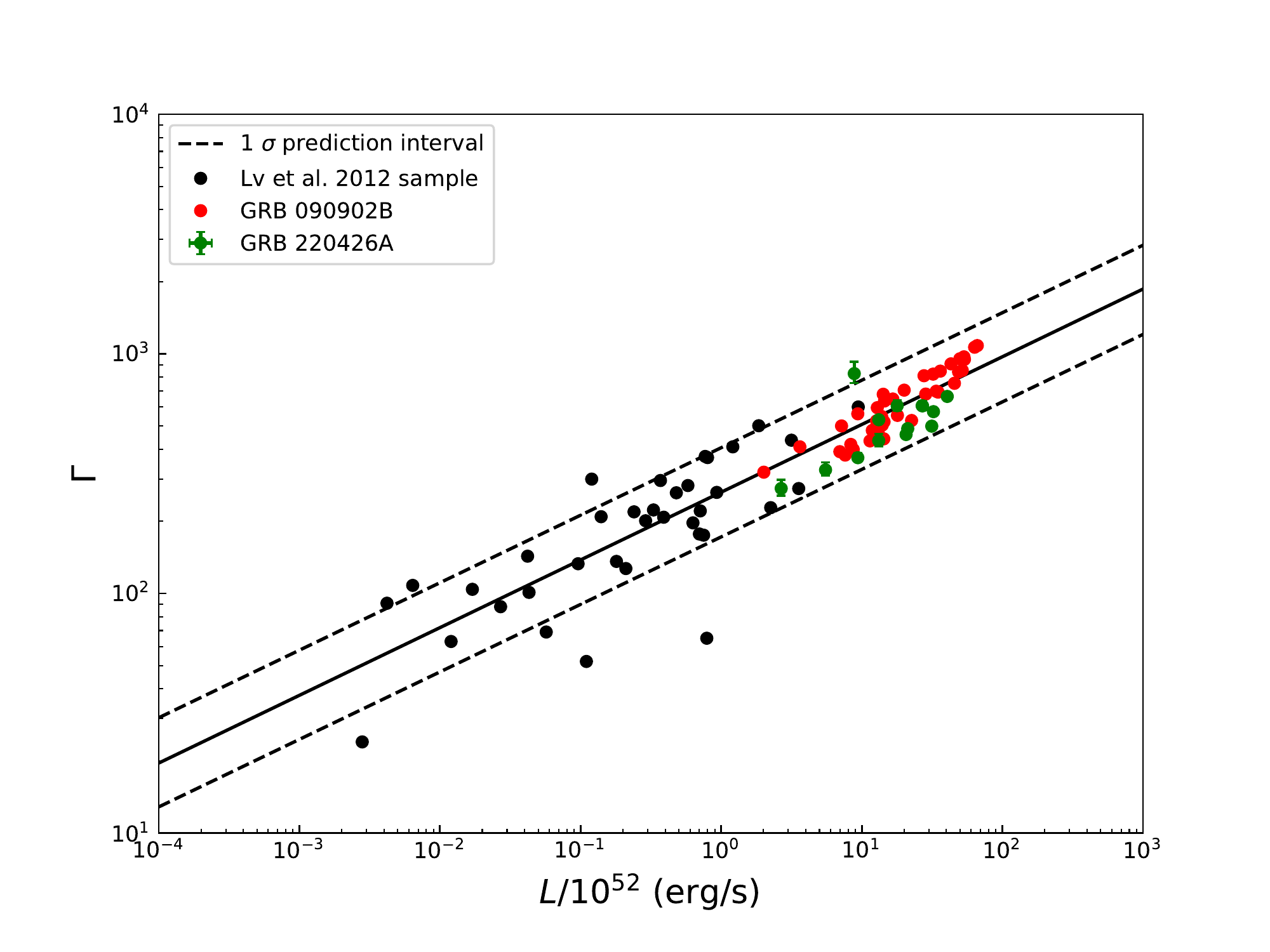}
\includegraphics[width=0.6\textwidth]{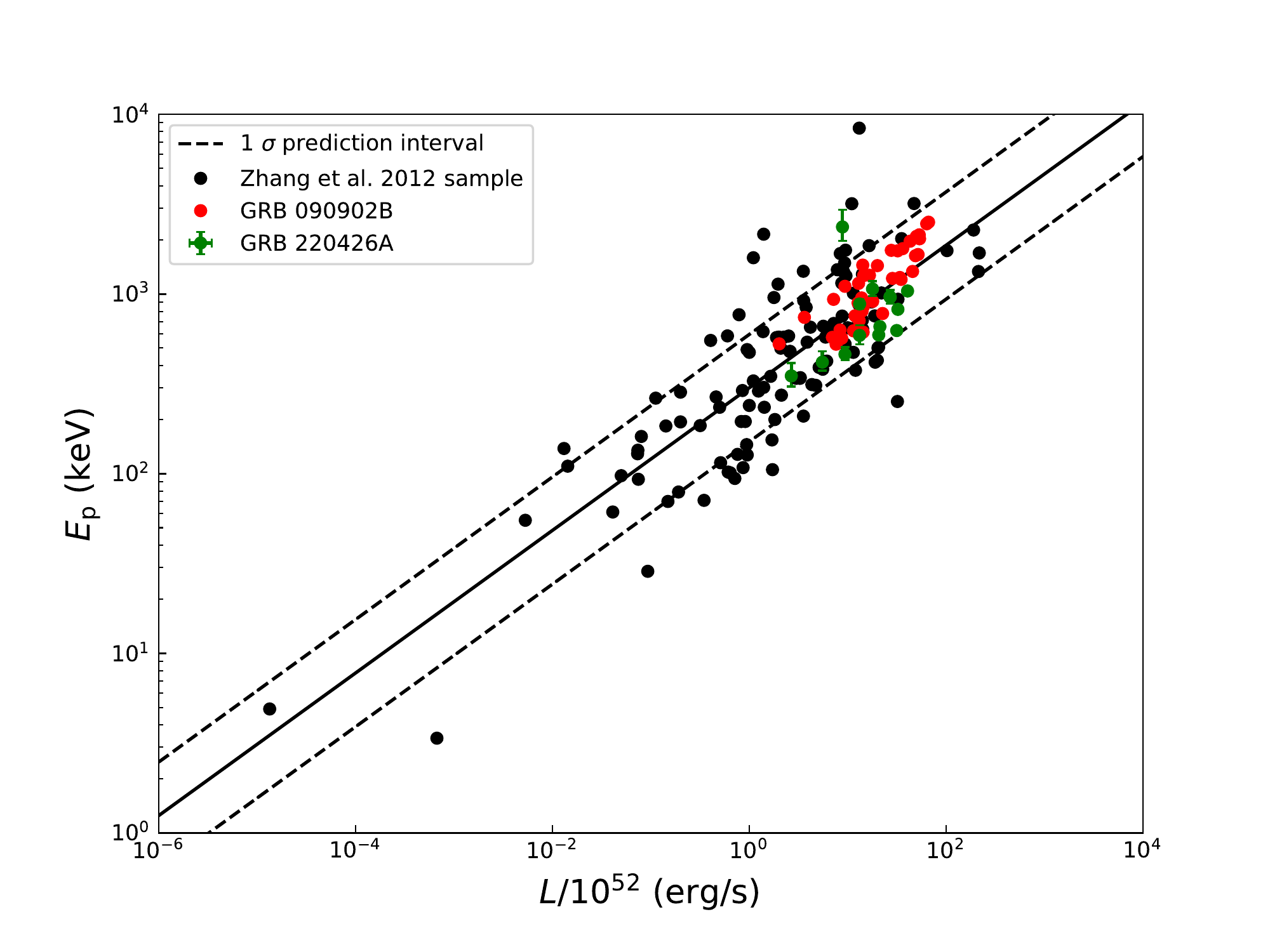}
\caption{Upper panel: $\Gamma-L$ diagram for for the bursts given by \cite{lu2012lorentz} (excluding those with a $\Gamma$ in dispute, for example GRB 090510 and GRB 090328A) and for the time-resolved thermal radiation of GRB 090902B and GRB 220426A.
	The black dashed line represents the fitting result of 1 $\sigma$ prediction interval,
	which is ${\rm log}\Gamma = 2.42_{-0.06}^{+0.06} +  0.28_{-0.06}^{+0.06}{\rm log}L $.
	Lower panel: $E_p-L$ diagram for the bursts investigated in \cite{zhang2012revisiting}.
	Consistent with the above, the black dashed line represents the fitting result of 1 $\sigma$ prediction interval,
	which is ${\rm log}E_{\rm p} = 2.47_{-0.05}^{+0.05} +  0.40_{-0.05}^{+0.04}{\rm log}L $.
}
\label{fig:fan_2012}
\end{figure}

\clearpage
\setlength{\tabcolsep}{2mm}{}
\begin{deluxetable}{ccccccccccccccccccc}
\label{tab:tab1}
\tabletypesize{\tiny}
\tablecaption{Time-integrated spectral fitting result}
\tablehead{\colhead{Time Interval} & \colhead{model}& \colhead{$\alpha$/$m$}&\colhead{$\beta$} &\colhead{$kT_{\rm min}$} &\colhead{$E_{p}$/$kT$/$kT_{\rm {max}}$} &\colhead{$\gamma$} &\colhead{Flux (8-40000 keV)} & $\ln{\cal Z}$ & {Favored Model}\\
	\colhead{(s)} &\colhead{}& \colhead{}&\colhead{} &\colhead{(keV)} &\colhead{(keV)} &\colhead{} &\colhead{($10^{-5}$ erg $\rm cm^{-2}$ $\rm s^{-1}$)} &\colhead{} &\colhead{}}
\startdata
[-0.05 - 7.79]&CPL  &-0.22$_{-0.01}^{+0.01}$ & {...} & ... &171.64$_{-0.80}^{+0.79}$ &... & 1.38 &-595.33  \\
&Band  &-0.13$_{-0.02}^{+0.02}$ & -3.56$_{-0.06}^{+0.06}$ & ... &162.73$_{-1.14}^{+1.15}$  &... & 1.46 &-516.67  \\
&BB  & ... & {...} & ... & 36.19$_{-0.10}^{+0.11}$ &... & 1.30 &-6991.28  \\
&mBB  & -0.18$_{-0.04}^{+0.04}$  & ... & 9.85$_{-0.23}^{+0.22}$ & 87.78$_{-1.15}^{+1.30}$ & ...  & 1.34 &-502.98  &  \\	
&CPL+PL  &-0.18$_{-0.02}^{+0.02}$ & {...} & ... &169.00$_{-0.91}^{+0.89}$ & -1.45$_{-0.05}^{+0.06}$ & 1.34+0.28&-586.15  \\
&Band+PL  &-0.13$_{-0.02}^{+0.02}$ & -3.56$_{-0.07}^{+0.06}$ &... &162.72$_{-1.11}^{+1.12}$ & -1.67$_{-0.58}^{+0.58}$ & ... &-517.02  \\
&BB+PL  & ... & {...} & ... &38.34$_{-0.16}^{+0.16}$ & -1.76$_{-0.01}^{+0.01}$ & 0.95+1.40 &-1959.90  \\
&mBB+PL & -0.36$_{-0.05}^{+0.05}$  & ... & 12.52$_{-0.33}^{+0.34}$ & 89.17$_{-1.61}^{+1.77}$ &  -1.90$_{-0.06}^{+0.04}$ & 1.29+0.21 &-442.53 &$\checkmark$ \\	
\enddata
\tablecomments{The calculation of Flux uses the median of the posterior distributions of the parameters in each model. 
	According to the model selection criterion given by Equation \ref{eq:7}, when an additional PL component is added to the model, except for the Band+PL model, the rest of the models get better goodness of fit ($\ln{\rm BF} > 8$).
	Therefore, we did not calculate the flux of the Band+PL model.}
\end{deluxetable}
\setlength{\tabcolsep}{2mm}{}
\begin{deluxetable}{ccccccccccccccccccc}
\label{tab:tab2}
\tabletypesize{}
\tablecaption{Time-resolved spectral fitting result}
\tablehead{\colhead{Time interval} & \colhead{model}& \colhead{$\alpha$/$m$}&\colhead{$\beta$} &\colhead{$kT_{\rm min}$} &\colhead{$E_{p}$/$kT$/$kT_{\rm max}$}  &\colhead{$\ln{\cal Z}$} & {favored model} \\
	\colhead{(s)} &\colhead{}& \colhead{}&\colhead{} &\colhead{(keV)} &\colhead{(keV)} &\colhead{} &\colhead{} & {}}
\startdata
[-0.05 - 0.62] & Band & 0.29$_{-0.11}^{+0.14}$ & -6.34$_{-2.48}^{+2.62}$ & ... & 338.20$_{-17.24}^{+13.95}$ & -263.35 \\ 
& CPL & 0.21$_{-0.09}^{+0.09}$ & ... & ... & 346.34$_{-12.05}^{+13.54}$ & -264.70 \\ 
& BB & ... & ... & ... & 73.43$_{-1.71}^{+1.68}$ & -295.46 \\ 
& mBB & -1.49$_{-0.16}^{+0.23}$ & ... & 41.27$_{-1.86}^{+1.61}$ & 250.81$_{-40.58}^{+62.12}$ & -258.08 &  $\checkmark$\\ 
\hline 
[0.62 - 0.88] & Band & 0.49$_{-0.10}^{+0.10}$ & -7.73$_{-1.56}^{+1.61}$ & ... & 204.51$_{-5.24}^{+5.30}$ & -279.24 \\ 
& CPL & 0.45$_{-0.09}^{+0.09}$ & ... & ... & 205.91$_{-5.36}^{+5.48}$ & -281.49 \\ 
& BB & ... & ... & ... & 48.13$_{-0.83}^{+0.85}$ & -309.16 \\ 
& mBB & -0.28$_{-0.15}^{+0.22}$ & ... & 21.88$_{-1.59}^{+1.53}$ & 93.47$_{-5.45}^{+5.89}$ & -272.68 &  $\checkmark$\\ 
\hline 
[0.88 - 1.33] & Band & 0.60$_{-0.11}^{+0.10}$ & -3.43$_{-0.28}^{+0.19}$ & ... & 168.11$_{-5.32}^{+5.68}$ & -271.08 \\ 
& CPL & 0.36$_{-0.06}^{+0.06}$ & ... & ... & 184.13$_{-2.94}^{+2.85}$ & -276.56 \\ 
& BB & ... & ... & ... & 42.64$_{-0.47}^{+0.47}$ & -387.66 \\ 
& mBB & -1.29$_{-0.20}^{+0.25}$ & ... & 22.15$_{-1.08}^{+0.86}$ & 113.37$_{-9.84}^{+11.99}$ & -259.66 &  $\checkmark$\\ 
\hline 
[1.33 - 2.22] & Band & 0.26$_{-0.04}^{+0.04}$ & -3.61$_{-0.17}^{+0.12}$ & ... & 196.25$_{-3.01}^{+2.84}$ & -351.22 \\ 
& CPL & 0.14$_{-0.03}^{+0.03}$ & ... & ... & 207.79$_{-2.02}^{+2.07}$ & -369.94 \\ 
& BB & ... & ... & ... & 46.23$_{-0.29}^{+0.30}$ & -950.79 \\ 
& mBB & -0.18$_{-0.11}^{+0.11}$ & ... & 16.09$_{-0.75}^{+0.66}$ & 102.46$_{-3.28}^{+3.21}$ & -335.69 &  $\checkmark$\\ 
\hline 
[2.22 - 2.61] & Band & 0.13$_{-0.04}^{+0.05}$ & -4.56$_{-1.35}^{+0.51}$ & ... & 229.20$_{-4.07}^{+3.88}$ & -249.29 \\ 
& CPL & 0.09$_{-0.03}^{+0.03}$ & ... & ... & 232.82$_{-2.90}^{+2.98}$ & -251.32 \\ 
& BB & ... & ... & ... & 51.21$_{-0.39}^{+0.42}$ & -675.97 \\ 
& mBB & -0.02$_{-0.05}^{+0.06}$ & ... & 16.08$_{-0.63}^{+0.58}$ & 110.41$_{-2.54}^{+2.54}$ & -229.09 &  $\checkmark$\\ 
\hline 
[2.61 - 2.99] & Band & 0.03$_{-0.04}^{+0.04}$ & -4.62$_{-2.04}^{+0.50}$ & ... & 204.63$_{-3.51}^{+3.53}$ & -277.08 \\ 
& CPL & 0.00$_{-0.04}^{+0.04}$ & ... & ... & 207.31$_{-2.65}^{+2.95}$ & -277.82 \\ 
& BB & ... & ... & ... & 45.50$_{-0.41}^{+0.40}$ & -681.78 \\ 
& mBB & 0.52$_{-0.11}^{+0.09}$ & ... & 8.32$_{-1.16}^{+1.17}$ & 87.25$_{-2.16}^{+2.71}$ & -276.13 &  $\checkmark$\\ 
\hline 
[2.99 - 3.92] & Band & 0.12$_{-0.04}^{+0.04}$ & -4.52$_{-0.75}^{+0.34}$ & ... & 148.83$_{-2.00}^{+2.10}$ & -348.68 \\ 
& CPL & 0.08$_{-0.03}^{+0.03}$ & ... & ... & 151.52$_{-1.42}^{+1.36}$ & -351.39 \\ 
& BB & ... & ... & ... & 34.46$_{-0.21}^{+0.21}$ & -988.03 \\ 
& mBB & 0.01$_{-0.15}^{+0.13}$ & ... & 11.01$_{-0.73}^{+0.78}$ & 69.94$_{-2.05}^{+2.59}$ & -336.78 &  $\checkmark$\\ 
\hline 
[3.92 - 4.14] & Band & 0.31$_{-0.11}^{+0.11}$ & -3.26$_{-0.22}^{+0.15}$ & ... & 130.61$_{-5.03}^{+5.67}$ & -247.56 \\ 
& CPL & -0.02$_{-0.06}^{+0.05}$ & ... & ... & 150.69$_{-2.59}^{+2.76}$ & -252.45 \\ 
& BB & ... & ... & ... & 33.07$_{-0.39}^{+0.41}$ & -476.59 \\ 
& mBB & -1.13$_{-0.17}^{+0.16}$ & ... & 14.72$_{-0.67}^{+0.74}$ & 101.99$_{-8.30}^{+9.94}$ & -232.57 &  $\checkmark$\\ 
\hline 
[4.14 - 4.75] & Band & 0.08$_{-0.04}^{+0.04}$ & -4.80$_{-0.88}^{+0.37}$ & ... & 135.38$_{-1.65}^{+1.62}$ & -291.27 \\ 
& CPL & 0.04$_{-0.03}^{+0.03}$ & ... & ... & 137.20$_{-1.20}^{+1.16}$ & -293.11 \\ 
& BB & ... & ... & ... & 31.40$_{-0.19}^{+0.18}$ & -1061.14 \\ 
& mBB & -0.24$_{-0.15}^{+0.12}$ & ... & 10.93$_{-0.59}^{+0.61}$ & 66.48$_{-1.98}^{+2.37}$ & -273.97 &  $\checkmark$\\ 
\hline 
[4.75 - 5.07] & Band & 0.02$_{-0.07}^{+0.08}$ & -3.76$_{-0.38}^{+0.26}$ & ... & 114.84$_{-3.48}^{+3.59}$ & -253.49 &  $\checkmark$\\ 
& CPL & -0.10$_{-0.05}^{+0.05}$ & ... & ... & 120.89$_{-1.90}^{+1.83}$ & -258.74 \\ 
& BB & ... & ... & ... & 26.71$_{-0.31}^{+0.31}$ & -551.66 \\ 
& mBB & -0.43$_{-0.21}^{+0.18}$ & ... & 9.58$_{-0.74}^{+0.77}$ & 62.84$_{-3.31}^{+4.42}$ & -254.32 \\ 
\hline 
[5.07 - 5.36] & Band & 0.07$_{-0.10}^{+0.11}$ & -4.77$_{-3.20}^{+0.82}$ & ... & 102.15$_{-3.75}^{+3.03}$ & -247.25 &  $\checkmark$\\ 
& CPL & -0.02$_{-0.07}^{+0.07}$ & ... & ... & 104.62$_{-2.08}^{+2.21}$ & -248.26 \\ 
& BB & ... & ... & ... & 23.13$_{-0.33}^{+0.34}$ & -362.61 \\ 
& mBB & -1.13$_{-0.36}^{+0.36}$ & ... & 11.34$_{-1.01}^{+0.92}$ & 62.49$_{-6.77}^{+8.80}$ & -249.72 \\ 
\hline 
[5.36 - 5.68] & Band & -0.21$_{-0.10}^{+0.13}$ & -4.89$_{-2.90}^{+1.23}$ & ... & 92.62$_{-4.24}^{+3.28}$ & -227.11  \\ 
& CPL & -0.29$_{-0.08}^{+0.08}$ & ... & ... & 94.69$_{-2.35}^{+2.60}$ & -226.88 & $\checkmark$ \\ 
& BB & ... & ... & ... & 20.21$_{-0.36}^{+0.33}$ & -346.96 \\ 
& mBB & -0.36$_{-0.29}^{+0.26}$ & ... & 6.90$_{-1.06}^{+1.01}$ & 49.09$_{-3.68}^{+4.66}$ & -231.44 \\ 
\hline 
[5.68 - 6.38] & Band & -0.24$_{-0.10}^{+0.11}$ & -3.60$_{-0.29}^{+0.24}$ & ... & 75.69$_{-2.27}^{+2.36}$ & -300.91 &  $\checkmark$\\ 
& CPL & -0.43$_{-0.07}^{+0.08}$ & ... & ... & 80.76$_{-1.99}^{+2.05}$ & -303.09 \\ 
& BB & ... & ... & ... & 17.37$_{-0.25}^{+0.25}$ & -444.69 \\ 
& mBB & -0.48$_{-0.37}^{+0.35}$ & ... & 5.90$_{-1.29}^{+1.10}$ & 44.35$_{-4.53}^{+6.62}$ & -308.74 \\ 
\hline 
[6.38 - 7.79] & Band & -0.32$_{-0.12}^{+0.13}$ & -3.51$_{-0.51}^{+0.26}$ & ... & 60.32$_{-2.12}^{+2.19}$ & -303.02 &  $\checkmark$\\ 
& CPL & -0.50$_{-0.09}^{+0.09}$ & ... & ... & 63.36$_{-1.67}^{+1.70}$ & -304.06 \\ 
& BB & ... & ... & ... & 14.32$_{-0.23}^{+0.24}$ & -421.76 \\ 
& mBB & -0.68$_{-0.36}^{+0.41}$ & ... & 5.22$_{-1.14}^{+0.83}$ & 37.17$_{-4.71}^{+6.72}$ & -306.69 \\ 
\enddata
\end{deluxetable}
\end{document}